\documentclass[12pt]{article}
\usepackage{epsfig,amsfonts,amssymb}
\usepackage{hyperref}

\usepackage{comment}
\usepackage{cite}
\usepackage{cite}

\def\ZZZ{{\hbox{ Z\kern-1.6mm Z}}}
\def\RRR{{\hbox{ R\kern-2.4mm R}}}
\def\CCC{{\hbox{ C\kern-2.0mm C}}}
\def\zzz{{\hbox{z\kern-1mm z}}}

\newcommand{\qeq}{{\hbox{=\kern-2.3mm ? \kern.5mm }}}
\renewcommand{\qeq}{=}

\newcommand{\eps}{\epsilon}

\newcommand{\DD}{{\cal D}}

\newcommand{\AAA}{{\cal A}}

\newcommand{\FF}{{\cal F}}
\newcommand{\JJ}{{\cal J}}

\newcommand{\OO}{{\cal O}}

\newcommand{\wt}{\widetilde}

\newcommand{\RR}{{\cal R}}
\newcommand{\NN}{{\cal N}}

\newcommand{\SSS}{{\cal S}}

\newcommand{\be}{\begin{equation}}
\newcommand{\ee}{\end{equation}}
\newcommand{\ben}{\begin{eqnarray}\displaystyle}
\newcommand{\een}{\end{eqnarray}}

\newcommand{\refb}[1]{(\ref{#1})}
\newcommand{\p}{\partial}
\newcommand{\sectiono}[1]{\section{#1}\setcounter{equation}{0}}

\def\one{{\hbox{ 1\kern-.8mm l}}}
\def\zero{{\hbox{ 0\kern-1.5mm 0}}}

\newcommand{\bea}[1]{\begin{eqnarray}\label{#1} }
\newcommand{\eea}{\end{eqnarray}}

\newcommand{\eqref}{\refb}

\usepackage{bm}
\usepackage[table]{xcolor}

\newcommand{\vea}{a}

\def\figone{

\def\JPicScale{0.4}
\ifx\JPicScale\undefined\def\JPicScale{1}\fi
\unitlength \JPicScale mm
\begin{picture}(140,90)(0,0)
\linethickness{0.1mm}
\put(30,60){\line(1,0){100}}
\linethickness{0.1mm}
\put(30,20){\line(1,0){100}}
\linethickness{0.3mm}
\multiput(30,90)(0.12,-0.18){167}{\line(0,-1){0.18}}
\linethickness{0.3mm}
\multiput(100,60)(0.12,0.12){250}{\line(1,0){0.12}}
\linethickness{0.3mm}
\multiput(100,20)(0.18,-0.12){167}{\line(1,0){0.18}}
\linethickness{0.3mm}
\multiput(30,0)(0.18,0.12){167}{\line(1,0){0.18}}
\put(150,62){\makebox(0,0)[cc]{$x^{\prime0}=T$}}

\put(155,22){\makebox(0,0)[cc]{$x^{\prime0}=-T$}}

\put(95,65){\makebox(0,0)[cc]{$c_{(1)}$}}

\put(140,90){\makebox(0,0)[cc]{$r_{(1)}$}}

\put(20,90){\makebox(0,0)[cc]{$r_{(2)}$}}

\put(140,0){\makebox(0,0)[cc]{$r_{(3)}$}}

\put(20,0){\makebox(0,0)[cc]{$r_{(4)}$}}

\put(55,65){\makebox(0,0)[cc]{$c_{(2)}$}}

\put(95,15){\makebox(0,0)[cc]{$c_{(3)}$}}

\put(65,15){\makebox(0,0)[cc]{$c_{(4)}$}}

\end{picture}

}

\def\figonea{

\def\JPicScale{0.5}
\ifx\JPicScale\undefined\def\JPicScale{1}\fi
\unitlength \JPicScale mm
\begin{picture}(135,90)(0,0)
\linethickness{0.3mm}
\multiput(45,90)(0.12,-0.18){167}{\line(0,-1){0.18}}
\linethickness{0.3mm}
\multiput(100,55)(0.12,0.12){250}{\line(1,0){0.12}}
\linethickness{0.3mm}
\multiput(100,25)(0.18,-0.12){167}{\line(1,0){0.18}}
\linethickness{0.3mm}
\multiput(30,5)(0.18,0.12){167}{\line(1,0){0.18}}
\put(110,55){\makebox(0,0)[cc]{$c_{(1)}$}}

\put(135,90){\makebox(0,0)[cc]{$R_{(1)}$}}

\put(35,90){\makebox(0,0)[cc]{$R_{(2)}$}}

\put(140,0){\makebox(0,0)[cc]{$R_{(3)}$}}

\put(20,0){\makebox(0,0)[cc]{$R_{(4)}$}}

\put(55,60){\makebox(0,0)[cc]{$c_{(2)}$}}

\put(110,26){\makebox(0,0)[cc]{$c_{(3)}$}}

\put(51,26){\makebox(0,0)[cc]{$c_{(4)}$}}

\linethickness{0.3mm}
\put(105.56,40.03){\line(0,1){0.5}}
\multiput(105.55,41.03)(0.01,-0.5){1}{\line(0,-1){0.5}}
\multiput(105.53,41.53)(0.02,-0.5){1}{\line(0,-1){0.5}}
\multiput(105.5,42.04)(0.03,-0.5){1}{\line(0,-1){0.5}}
\multiput(105.46,42.54)(0.04,-0.5){1}{\line(0,-1){0.5}}
\multiput(105.41,43.04)(0.05,-0.5){1}{\line(0,-1){0.5}}
\multiput(105.35,43.54)(0.06,-0.5){1}{\line(0,-1){0.5}}
\multiput(105.28,44.03)(0.07,-0.5){1}{\line(0,-1){0.5}}
\multiput(105.2,44.53)(0.08,-0.5){1}{\line(0,-1){0.5}}
\multiput(105.11,45.02)(0.09,-0.49){1}{\line(0,-1){0.49}}
\multiput(105.01,45.52)(0.1,-0.49){1}{\line(0,-1){0.49}}
\multiput(104.9,46.01)(0.11,-0.49){1}{\line(0,-1){0.49}}
\multiput(104.78,46.5)(0.12,-0.49){1}{\line(0,-1){0.49}}
\multiput(104.65,46.98)(0.13,-0.49){1}{\line(0,-1){0.49}}
\multiput(104.51,47.47)(0.14,-0.48){1}{\line(0,-1){0.48}}
\multiput(104.37,47.95)(0.15,-0.48){1}{\line(0,-1){0.48}}
\multiput(104.21,48.42)(0.16,-0.48){1}{\line(0,-1){0.48}}
\multiput(104.04,48.9)(0.17,-0.47){1}{\line(0,-1){0.47}}
\multiput(103.87,49.37)(0.18,-0.47){1}{\line(0,-1){0.47}}
\multiput(103.68,49.84)(0.09,-0.23){2}{\line(0,-1){0.23}}
\multiput(103.49,50.3)(0.1,-0.23){2}{\line(0,-1){0.23}}
\multiput(103.28,50.76)(0.1,-0.23){2}{\line(0,-1){0.23}}
\multiput(103.07,51.21)(0.11,-0.23){2}{\line(0,-1){0.23}}
\multiput(102.85,51.66)(0.11,-0.23){2}{\line(0,-1){0.23}}
\multiput(102.62,52.11)(0.12,-0.22){2}{\line(0,-1){0.22}}
\multiput(102.38,52.55)(0.12,-0.22){2}{\line(0,-1){0.22}}
\multiput(102.13,52.99)(0.12,-0.22){2}{\line(0,-1){0.22}}
\multiput(101.87,53.42)(0.13,-0.22){2}{\line(0,-1){0.22}}
\multiput(101.61,53.85)(0.13,-0.21){2}{\line(0,-1){0.21}}
\multiput(101.33,54.27)(0.14,-0.21){2}{\line(0,-1){0.21}}
\multiput(101.05,54.69)(0.14,-0.21){2}{\line(0,-1){0.21}}
\multiput(100.76,55.1)(0.15,-0.21){2}{\line(0,-1){0.21}}
\multiput(100.46,55.5)(0.15,-0.2){2}{\line(0,-1){0.2}}
\multiput(100.15,55.9)(0.1,-0.13){3}{\line(0,-1){0.13}}
\multiput(99.84,56.29)(0.1,-0.13){3}{\line(0,-1){0.13}}
\multiput(99.52,56.68)(0.11,-0.13){3}{\line(0,-1){0.13}}
\multiput(99.19,57.06)(0.11,-0.13){3}{\line(0,-1){0.13}}
\multiput(98.85,57.43)(0.11,-0.12){3}{\line(0,-1){0.12}}
\multiput(98.5,57.79)(0.11,-0.12){3}{\line(0,-1){0.12}}
\multiput(98.15,58.15)(0.12,-0.12){3}{\line(0,-1){0.12}}
\multiput(97.79,58.5)(0.12,-0.12){3}{\line(1,0){0.12}}
\multiput(97.43,58.85)(0.12,-0.11){3}{\line(1,0){0.12}}
\multiput(97.06,59.19)(0.12,-0.11){3}{\line(1,0){0.12}}
\multiput(96.68,59.52)(0.13,-0.11){3}{\line(1,0){0.13}}
\multiput(96.29,59.84)(0.13,-0.11){3}{\line(1,0){0.13}}
\multiput(95.9,60.15)(0.13,-0.1){3}{\line(1,0){0.13}}
\multiput(95.5,60.46)(0.13,-0.1){3}{\line(1,0){0.13}}
\multiput(95.1,60.76)(0.2,-0.15){2}{\line(1,0){0.2}}
\multiput(94.69,61.05)(0.21,-0.15){2}{\line(1,0){0.21}}
\multiput(94.27,61.33)(0.21,-0.14){2}{\line(1,0){0.21}}
\multiput(93.85,61.61)(0.21,-0.14){2}{\line(1,0){0.21}}
\multiput(93.42,61.87)(0.21,-0.13){2}{\line(1,0){0.21}}
\multiput(92.99,62.13)(0.22,-0.13){2}{\line(1,0){0.22}}
\multiput(92.55,62.38)(0.22,-0.12){2}{\line(1,0){0.22}}
\multiput(92.11,62.62)(0.22,-0.12){2}{\line(1,0){0.22}}
\multiput(91.66,62.85)(0.22,-0.12){2}{\line(1,0){0.22}}
\multiput(91.21,63.07)(0.23,-0.11){2}{\line(1,0){0.23}}
\multiput(90.76,63.28)(0.23,-0.11){2}{\line(1,0){0.23}}
\multiput(90.3,63.49)(0.23,-0.1){2}{\line(1,0){0.23}}
\multiput(89.84,63.68)(0.23,-0.1){2}{\line(1,0){0.23}}
\multiput(89.37,63.87)(0.23,-0.09){2}{\line(1,0){0.23}}
\multiput(88.9,64.04)(0.47,-0.18){1}{\line(1,0){0.47}}
\multiput(88.42,64.21)(0.47,-0.17){1}{\line(1,0){0.47}}
\multiput(87.95,64.37)(0.48,-0.16){1}{\line(1,0){0.48}}
\multiput(87.47,64.51)(0.48,-0.15){1}{\line(1,0){0.48}}
\multiput(86.98,64.65)(0.48,-0.14){1}{\line(1,0){0.48}}
\multiput(86.5,64.78)(0.49,-0.13){1}{\line(1,0){0.49}}
\multiput(86.01,64.9)(0.49,-0.12){1}{\line(1,0){0.49}}
\multiput(85.52,65.01)(0.49,-0.11){1}{\line(1,0){0.49}}
\multiput(85.02,65.11)(0.49,-0.1){1}{\line(1,0){0.49}}
\multiput(84.53,65.2)(0.49,-0.09){1}{\line(1,0){0.49}}
\multiput(84.03,65.28)(0.5,-0.08){1}{\line(1,0){0.5}}
\multiput(83.54,65.35)(0.5,-0.07){1}{\line(1,0){0.5}}
\multiput(83.04,65.41)(0.5,-0.06){1}{\line(1,0){0.5}}
\multiput(82.54,65.46)(0.5,-0.05){1}{\line(1,0){0.5}}
\multiput(82.04,65.5)(0.5,-0.04){1}{\line(1,0){0.5}}
\multiput(81.53,65.53)(0.5,-0.03){1}{\line(1,0){0.5}}
\multiput(81.03,65.55)(0.5,-0.02){1}{\line(1,0){0.5}}
\multiput(80.53,65.56)(0.5,-0.01){1}{\line(1,0){0.5}}
\put(80.03,65.56){\line(1,0){0.5}}
\multiput(79.52,65.55)(0.5,0.01){1}{\line(1,0){0.5}}
\multiput(79.02,65.53)(0.5,0.02){1}{\line(1,0){0.5}}
\multiput(78.52,65.5)(0.5,0.03){1}{\line(1,0){0.5}}
\multiput(78.02,65.46)(0.5,0.04){1}{\line(1,0){0.5}}
\multiput(77.52,65.41)(0.5,0.05){1}{\line(1,0){0.5}}
\multiput(77.02,65.35)(0.5,0.06){1}{\line(1,0){0.5}}
\multiput(76.52,65.28)(0.5,0.07){1}{\line(1,0){0.5}}
\multiput(76.03,65.2)(0.5,0.08){1}{\line(1,0){0.5}}
\multiput(75.53,65.11)(0.49,0.09){1}{\line(1,0){0.49}}
\multiput(75.04,65.01)(0.49,0.1){1}{\line(1,0){0.49}}
\multiput(74.55,64.9)(0.49,0.11){1}{\line(1,0){0.49}}
\multiput(74.06,64.78)(0.49,0.12){1}{\line(1,0){0.49}}
\multiput(73.57,64.65)(0.49,0.13){1}{\line(1,0){0.49}}
\multiput(73.09,64.51)(0.48,0.14){1}{\line(1,0){0.48}}
\multiput(72.61,64.37)(0.48,0.15){1}{\line(1,0){0.48}}
\multiput(72.13,64.21)(0.48,0.16){1}{\line(1,0){0.48}}
\multiput(71.66,64.04)(0.47,0.17){1}{\line(1,0){0.47}}
\multiput(71.19,63.87)(0.47,0.18){1}{\line(1,0){0.47}}
\multiput(70.72,63.68)(0.23,0.09){2}{\line(1,0){0.23}}
\multiput(70.26,63.49)(0.23,0.1){2}{\line(1,0){0.23}}
\multiput(69.8,63.28)(0.23,0.1){2}{\line(1,0){0.23}}
\multiput(69.34,63.07)(0.23,0.11){2}{\line(1,0){0.23}}
\multiput(68.89,62.85)(0.23,0.11){2}{\line(1,0){0.23}}
\multiput(68.44,62.62)(0.22,0.12){2}{\line(1,0){0.22}}
\multiput(68,62.38)(0.22,0.12){2}{\line(1,0){0.22}}
\multiput(67.57,62.13)(0.22,0.12){2}{\line(1,0){0.22}}
\multiput(67.13,61.87)(0.22,0.13){2}{\line(1,0){0.22}}
\multiput(66.71,61.61)(0.21,0.13){2}{\line(1,0){0.21}}
\multiput(66.29,61.33)(0.21,0.14){2}{\line(1,0){0.21}}
\multiput(65.87,61.05)(0.21,0.14){2}{\line(1,0){0.21}}
\multiput(65.46,60.76)(0.21,0.15){2}{\line(1,0){0.21}}
\multiput(65.06,60.46)(0.2,0.15){2}{\line(1,0){0.2}}
\multiput(64.66,60.15)(0.13,0.1){3}{\line(1,0){0.13}}
\multiput(64.27,59.84)(0.13,0.1){3}{\line(1,0){0.13}}
\multiput(63.88,59.52)(0.13,0.11){3}{\line(1,0){0.13}}
\multiput(63.5,59.19)(0.13,0.11){3}{\line(1,0){0.13}}
\multiput(63.13,58.85)(0.12,0.11){3}{\line(1,0){0.12}}
\multiput(62.76,58.5)(0.12,0.11){3}{\line(1,0){0.12}}
\multiput(62.4,58.15)(0.12,0.12){3}{\line(1,0){0.12}}
\multiput(62.05,57.79)(0.12,0.12){3}{\line(0,1){0.12}}
\multiput(61.71,57.43)(0.11,0.12){3}{\line(0,1){0.12}}
\multiput(61.37,57.06)(0.11,0.12){3}{\line(0,1){0.12}}
\multiput(61.04,56.68)(0.11,0.13){3}{\line(0,1){0.13}}
\multiput(60.72,56.29)(0.11,0.13){3}{\line(0,1){0.13}}
\multiput(60.4,55.9)(0.1,0.13){3}{\line(0,1){0.13}}
\multiput(60.1,55.5)(0.1,0.13){3}{\line(0,1){0.13}}
\multiput(59.8,55.1)(0.15,0.2){2}{\line(0,1){0.2}}
\multiput(59.51,54.69)(0.15,0.21){2}{\line(0,1){0.21}}
\multiput(59.22,54.27)(0.14,0.21){2}{\line(0,1){0.21}}
\multiput(58.95,53.85)(0.14,0.21){2}{\line(0,1){0.21}}
\multiput(58.68,53.42)(0.13,0.21){2}{\line(0,1){0.21}}
\multiput(58.43,52.99)(0.13,0.22){2}{\line(0,1){0.22}}
\multiput(58.18,52.55)(0.12,0.22){2}{\line(0,1){0.22}}
\multiput(57.94,52.11)(0.12,0.22){2}{\line(0,1){0.22}}
\multiput(57.71,51.66)(0.12,0.22){2}{\line(0,1){0.22}}
\multiput(57.49,51.21)(0.11,0.23){2}{\line(0,1){0.23}}
\multiput(57.27,50.76)(0.11,0.23){2}{\line(0,1){0.23}}
\multiput(57.07,50.3)(0.1,0.23){2}{\line(0,1){0.23}}
\multiput(56.87,49.84)(0.1,0.23){2}{\line(0,1){0.23}}
\multiput(56.69,49.37)(0.09,0.23){2}{\line(0,1){0.23}}
\multiput(56.51,48.9)(0.18,0.47){1}{\line(0,1){0.47}}
\multiput(56.35,48.42)(0.17,0.47){1}{\line(0,1){0.47}}
\multiput(56.19,47.95)(0.16,0.48){1}{\line(0,1){0.48}}
\multiput(56.04,47.47)(0.15,0.48){1}{\line(0,1){0.48}}
\multiput(55.9,46.98)(0.14,0.48){1}{\line(0,1){0.48}}
\multiput(55.78,46.5)(0.13,0.49){1}{\line(0,1){0.49}}
\multiput(55.66,46.01)(0.12,0.49){1}{\line(0,1){0.49}}
\multiput(55.55,45.52)(0.11,0.49){1}{\line(0,1){0.49}}
\multiput(55.45,45.02)(0.1,0.49){1}{\line(0,1){0.49}}
\multiput(55.36,44.53)(0.09,0.49){1}{\line(0,1){0.49}}
\multiput(55.28,44.03)(0.08,0.5){1}{\line(0,1){0.5}}
\multiput(55.21,43.54)(0.07,0.5){1}{\line(0,1){0.5}}
\multiput(55.15,43.04)(0.06,0.5){1}{\line(0,1){0.5}}
\multiput(55.1,42.54)(0.05,0.5){1}{\line(0,1){0.5}}
\multiput(55.06,42.04)(0.04,0.5){1}{\line(0,1){0.5}}
\multiput(55.03,41.53)(0.03,0.5){1}{\line(0,1){0.5}}
\multiput(55.01,41.03)(0.02,0.5){1}{\line(0,1){0.5}}
\multiput(55,40.53)(0.01,0.5){1}{\line(0,1){0.5}}
\put(55,40.03){\line(0,1){0.5}}
\multiput(55,40.03)(0.01,-0.5){1}{\line(0,-1){0.5}}
\multiput(55.01,39.52)(0.02,-0.5){1}{\line(0,-1){0.5}}
\multiput(55.03,39.02)(0.03,-0.5){1}{\line(0,-1){0.5}}
\multiput(55.06,38.52)(0.04,-0.5){1}{\line(0,-1){0.5}}
\multiput(55.1,38.02)(0.05,-0.5){1}{\line(0,-1){0.5}}
\multiput(55.15,37.52)(0.06,-0.5){1}{\line(0,-1){0.5}}
\multiput(55.21,37.02)(0.07,-0.5){1}{\line(0,-1){0.5}}
\multiput(55.28,36.52)(0.08,-0.5){1}{\line(0,-1){0.5}}
\multiput(55.36,36.03)(0.09,-0.49){1}{\line(0,-1){0.49}}
\multiput(55.45,35.53)(0.1,-0.49){1}{\line(0,-1){0.49}}
\multiput(55.55,35.04)(0.11,-0.49){1}{\line(0,-1){0.49}}
\multiput(55.66,34.55)(0.12,-0.49){1}{\line(0,-1){0.49}}
\multiput(55.78,34.06)(0.13,-0.49){1}{\line(0,-1){0.49}}
\multiput(55.9,33.57)(0.14,-0.48){1}{\line(0,-1){0.48}}
\multiput(56.04,33.09)(0.15,-0.48){1}{\line(0,-1){0.48}}
\multiput(56.19,32.61)(0.16,-0.48){1}{\line(0,-1){0.48}}
\multiput(56.35,32.13)(0.17,-0.47){1}{\line(0,-1){0.47}}
\multiput(56.51,31.66)(0.18,-0.47){1}{\line(0,-1){0.47}}
\multiput(56.69,31.19)(0.09,-0.23){2}{\line(0,-1){0.23}}
\multiput(56.87,30.72)(0.1,-0.23){2}{\line(0,-1){0.23}}
\multiput(57.07,30.26)(0.1,-0.23){2}{\line(0,-1){0.23}}
\multiput(57.27,29.8)(0.11,-0.23){2}{\line(0,-1){0.23}}
\multiput(57.49,29.34)(0.11,-0.23){2}{\line(0,-1){0.23}}
\multiput(57.71,28.89)(0.12,-0.22){2}{\line(0,-1){0.22}}
\multiput(57.94,28.44)(0.12,-0.22){2}{\line(0,-1){0.22}}
\multiput(58.18,28)(0.12,-0.22){2}{\line(0,-1){0.22}}
\multiput(58.43,27.57)(0.13,-0.22){2}{\line(0,-1){0.22}}
\multiput(58.68,27.13)(0.13,-0.21){2}{\line(0,-1){0.21}}
\multiput(58.95,26.71)(0.14,-0.21){2}{\line(0,-1){0.21}}
\multiput(59.22,26.29)(0.14,-0.21){2}{\line(0,-1){0.21}}
\multiput(59.51,25.87)(0.15,-0.21){2}{\line(0,-1){0.21}}
\multiput(59.8,25.46)(0.15,-0.2){2}{\line(0,-1){0.2}}
\multiput(60.1,25.06)(0.1,-0.13){3}{\line(0,-1){0.13}}
\multiput(60.4,24.66)(0.1,-0.13){3}{\line(0,-1){0.13}}
\multiput(60.72,24.27)(0.11,-0.13){3}{\line(0,-1){0.13}}
\multiput(61.04,23.88)(0.11,-0.13){3}{\line(0,-1){0.13}}
\multiput(61.37,23.5)(0.11,-0.12){3}{\line(0,-1){0.12}}
\multiput(61.71,23.13)(0.11,-0.12){3}{\line(0,-1){0.12}}
\multiput(62.05,22.76)(0.12,-0.12){3}{\line(0,-1){0.12}}
\multiput(62.4,22.4)(0.12,-0.12){3}{\line(1,0){0.12}}
\multiput(62.76,22.05)(0.12,-0.11){3}{\line(1,0){0.12}}
\multiput(63.13,21.71)(0.12,-0.11){3}{\line(1,0){0.12}}
\multiput(63.5,21.37)(0.13,-0.11){3}{\line(1,0){0.13}}
\multiput(63.88,21.04)(0.13,-0.11){3}{\line(1,0){0.13}}
\multiput(64.27,20.72)(0.13,-0.1){3}{\line(1,0){0.13}}
\multiput(64.66,20.4)(0.13,-0.1){3}{\line(1,0){0.13}}
\multiput(65.06,20.1)(0.2,-0.15){2}{\line(1,0){0.2}}
\multiput(65.46,19.8)(0.21,-0.15){2}{\line(1,0){0.21}}
\multiput(65.87,19.51)(0.21,-0.14){2}{\line(1,0){0.21}}
\multiput(66.29,19.22)(0.21,-0.14){2}{\line(1,0){0.21}}
\multiput(66.71,18.95)(0.21,-0.13){2}{\line(1,0){0.21}}
\multiput(67.13,18.68)(0.22,-0.13){2}{\line(1,0){0.22}}
\multiput(67.57,18.43)(0.22,-0.12){2}{\line(1,0){0.22}}
\multiput(68,18.18)(0.22,-0.12){2}{\line(1,0){0.22}}
\multiput(68.44,17.94)(0.22,-0.12){2}{\line(1,0){0.22}}
\multiput(68.89,17.71)(0.23,-0.11){2}{\line(1,0){0.23}}
\multiput(69.34,17.49)(0.23,-0.11){2}{\line(1,0){0.23}}
\multiput(69.8,17.27)(0.23,-0.1){2}{\line(1,0){0.23}}
\multiput(70.26,17.07)(0.23,-0.1){2}{\line(1,0){0.23}}
\multiput(70.72,16.87)(0.23,-0.09){2}{\line(1,0){0.23}}
\multiput(71.19,16.69)(0.47,-0.18){1}{\line(1,0){0.47}}
\multiput(71.66,16.51)(0.47,-0.17){1}{\line(1,0){0.47}}
\multiput(72.13,16.35)(0.48,-0.16){1}{\line(1,0){0.48}}
\multiput(72.61,16.19)(0.48,-0.15){1}{\line(1,0){0.48}}
\multiput(73.09,16.04)(0.48,-0.14){1}{\line(1,0){0.48}}
\multiput(73.57,15.9)(0.49,-0.13){1}{\line(1,0){0.49}}
\multiput(74.06,15.78)(0.49,-0.12){1}{\line(1,0){0.49}}
\multiput(74.55,15.66)(0.49,-0.11){1}{\line(1,0){0.49}}
\multiput(75.04,15.55)(0.49,-0.1){1}{\line(1,0){0.49}}
\multiput(75.53,15.45)(0.49,-0.09){1}{\line(1,0){0.49}}
\multiput(76.03,15.36)(0.5,-0.08){1}{\line(1,0){0.5}}
\multiput(76.52,15.28)(0.5,-0.07){1}{\line(1,0){0.5}}
\multiput(77.02,15.21)(0.5,-0.06){1}{\line(1,0){0.5}}
\multiput(77.52,15.15)(0.5,-0.05){1}{\line(1,0){0.5}}
\multiput(78.02,15.1)(0.5,-0.04){1}{\line(1,0){0.5}}
\multiput(78.52,15.06)(0.5,-0.03){1}{\line(1,0){0.5}}
\multiput(79.02,15.03)(0.5,-0.02){1}{\line(1,0){0.5}}
\multiput(79.52,15.01)(0.5,-0.01){1}{\line(1,0){0.5}}
\put(80.03,15){\line(1,0){0.5}}
\multiput(80.53,15)(0.5,0.01){1}{\line(1,0){0.5}}
\multiput(81.03,15.01)(0.5,0.02){1}{\line(1,0){0.5}}
\multiput(81.53,15.03)(0.5,0.03){1}{\line(1,0){0.5}}
\multiput(82.04,15.06)(0.5,0.04){1}{\line(1,0){0.5}}
\multiput(82.54,15.1)(0.5,0.05){1}{\line(1,0){0.5}}
\multiput(83.04,15.15)(0.5,0.06){1}{\line(1,0){0.5}}
\multiput(83.54,15.21)(0.5,0.07){1}{\line(1,0){0.5}}
\multiput(84.03,15.28)(0.5,0.08){1}{\line(1,0){0.5}}
\multiput(84.53,15.36)(0.49,0.09){1}{\line(1,0){0.49}}
\multiput(85.02,15.45)(0.49,0.1){1}{\line(1,0){0.49}}
\multiput(85.52,15.55)(0.49,0.11){1}{\line(1,0){0.49}}
\multiput(86.01,15.66)(0.49,0.12){1}{\line(1,0){0.49}}
\multiput(86.5,15.78)(0.49,0.13){1}{\line(1,0){0.49}}
\multiput(86.98,15.9)(0.48,0.14){1}{\line(1,0){0.48}}
\multiput(87.47,16.04)(0.48,0.15){1}{\line(1,0){0.48}}
\multiput(87.95,16.19)(0.48,0.16){1}{\line(1,0){0.48}}
\multiput(88.42,16.35)(0.47,0.17){1}{\line(1,0){0.47}}
\multiput(88.9,16.51)(0.47,0.18){1}{\line(1,0){0.47}}
\multiput(89.37,16.69)(0.23,0.09){2}{\line(1,0){0.23}}
\multiput(89.84,16.87)(0.23,0.1){2}{\line(1,0){0.23}}
\multiput(90.3,17.07)(0.23,0.1){2}{\line(1,0){0.23}}
\multiput(90.76,17.27)(0.23,0.11){2}{\line(1,0){0.23}}
\multiput(91.21,17.49)(0.23,0.11){2}{\line(1,0){0.23}}
\multiput(91.66,17.71)(0.22,0.12){2}{\line(1,0){0.22}}
\multiput(92.11,17.94)(0.22,0.12){2}{\line(1,0){0.22}}
\multiput(92.55,18.18)(0.22,0.12){2}{\line(1,0){0.22}}
\multiput(92.99,18.43)(0.22,0.13){2}{\line(1,0){0.22}}
\multiput(93.42,18.68)(0.21,0.13){2}{\line(1,0){0.21}}
\multiput(93.85,18.95)(0.21,0.14){2}{\line(1,0){0.21}}
\multiput(94.27,19.22)(0.21,0.14){2}{\line(1,0){0.21}}
\multiput(94.69,19.51)(0.21,0.15){2}{\line(1,0){0.21}}
\multiput(95.1,19.8)(0.2,0.15){2}{\line(1,0){0.2}}
\multiput(95.5,20.1)(0.13,0.1){3}{\line(1,0){0.13}}
\multiput(95.9,20.4)(0.13,0.1){3}{\line(1,0){0.13}}
\multiput(96.29,20.72)(0.13,0.11){3}{\line(1,0){0.13}}
\multiput(96.68,21.04)(0.13,0.11){3}{\line(1,0){0.13}}
\multiput(97.06,21.37)(0.12,0.11){3}{\line(1,0){0.12}}
\multiput(97.43,21.71)(0.12,0.11){3}{\line(1,0){0.12}}
\multiput(97.79,22.05)(0.12,0.12){3}{\line(1,0){0.12}}
\multiput(98.15,22.4)(0.12,0.12){3}{\line(0,1){0.12}}
\multiput(98.5,22.76)(0.11,0.12){3}{\line(0,1){0.12}}
\multiput(98.85,23.13)(0.11,0.12){3}{\line(0,1){0.12}}
\multiput(99.19,23.5)(0.11,0.13){3}{\line(0,1){0.13}}
\multiput(99.52,23.88)(0.11,0.13){3}{\line(0,1){0.13}}
\multiput(99.84,24.27)(0.1,0.13){3}{\line(0,1){0.13}}
\multiput(100.15,24.66)(0.1,0.13){3}{\line(0,1){0.13}}
\multiput(100.46,25.06)(0.15,0.2){2}{\line(0,1){0.2}}
\multiput(100.76,25.46)(0.15,0.21){2}{\line(0,1){0.21}}
\multiput(101.05,25.87)(0.14,0.21){2}{\line(0,1){0.21}}
\multiput(101.33,26.29)(0.14,0.21){2}{\line(0,1){0.21}}
\multiput(101.61,26.71)(0.13,0.21){2}{\line(0,1){0.21}}
\multiput(101.87,27.13)(0.13,0.22){2}{\line(0,1){0.22}}
\multiput(102.13,27.57)(0.12,0.22){2}{\line(0,1){0.22}}
\multiput(102.38,28)(0.12,0.22){2}{\line(0,1){0.22}}
\multiput(102.62,28.44)(0.12,0.22){2}{\line(0,1){0.22}}
\multiput(102.85,28.89)(0.11,0.23){2}{\line(0,1){0.23}}
\multiput(103.07,29.34)(0.11,0.23){2}{\line(0,1){0.23}}
\multiput(103.28,29.8)(0.1,0.23){2}{\line(0,1){0.23}}
\multiput(103.49,30.26)(0.1,0.23){2}{\line(0,1){0.23}}
\multiput(103.68,30.72)(0.09,0.23){2}{\line(0,1){0.23}}
\multiput(103.87,31.19)(0.18,0.47){1}{\line(0,1){0.47}}
\multiput(104.04,31.66)(0.17,0.47){1}{\line(0,1){0.47}}
\multiput(104.21,32.13)(0.16,0.48){1}{\line(0,1){0.48}}
\multiput(104.37,32.61)(0.15,0.48){1}{\line(0,1){0.48}}
\multiput(104.51,33.09)(0.14,0.48){1}{\line(0,1){0.48}}
\multiput(104.65,33.57)(0.13,0.49){1}{\line(0,1){0.49}}
\multiput(104.78,34.06)(0.12,0.49){1}{\line(0,1){0.49}}
\multiput(104.9,34.55)(0.11,0.49){1}{\line(0,1){0.49}}
\multiput(105.01,35.04)(0.1,0.49){1}{\line(0,1){0.49}}
\multiput(105.11,35.53)(0.09,0.49){1}{\line(0,1){0.49}}
\multiput(105.2,36.03)(0.08,0.5){1}{\line(0,1){0.5}}
\multiput(105.28,36.52)(0.07,0.5){1}{\line(0,1){0.5}}
\multiput(105.35,37.02)(0.06,0.5){1}{\line(0,1){0.5}}
\multiput(105.41,37.52)(0.05,0.5){1}{\line(0,1){0.5}}
\multiput(105.46,38.02)(0.04,0.5){1}{\line(0,1){0.5}}
\multiput(105.5,38.52)(0.03,0.5){1}{\line(0,1){0.5}}
\multiput(105.53,39.02)(0.02,0.5){1}{\line(0,1){0.5}}
\multiput(105.55,39.52)(0.01,0.5){1}{\line(0,1){0.5}}

\end{picture}

}

\def\tcr{\textcolor{red}}

\begin{document}

\baselineskip 24pt

\begin{center}

{\Large \bf A Classical Proof of the Classical Soft Graviton Theorem in D$>$4}


\end{center}

\vskip .6cm
\medskip

\vspace*{4.0ex}

\baselineskip=18pt

\centerline{\large \rm Alok Laddha$^{a}$ and Ashoke Sen$^{b}$}

\vspace*{4.0ex}

\centerline{\large \it ~$^a$Chennai Mathematical Institute, Siruseri, Chennai, India}

\centerline{\large \it ~$^b$Harish-Chandra Research Institute, HBNI}
\centerline{\large \it  Chhatnag Road, Jhusi,
Allahabad 211019, India}


\vspace*{1.0ex}
\centerline{\small E-mail:  aladdha@cmi.ac.in, sen@hri.res.in}

\vspace*{5.0ex}

\centerline{\bf Abstract} \bigskip

Classical soft graviton theorem gives an expression for the spectrum of
low frequency gravitational radiation, emitted during 
a classical scattering process, in terms of the
trajectories and spin angular momenta of ingoing and outgoing objects, including hard radiation. 
This has been proved to subleading order in the expansion
in powers of frequency by taking the classical limit of the quantum soft graviton theorem. In this paper we give
a direct proof of this result by analyzing the
classical equations of motion of a generic theory of gravity coupled to interacting matter in space-time
dimensions larger than four. 

\vfill \eject

\baselineskip 18pt

\tableofcontents

\sectiono{Introduction} \label{s0}

Classical soft graviton theorem\cite{1801.07719} 
describes the spectrum of low frequency gravitational radiation emitted during a 
classical scattering process, including decay in which a single bound system explodes into a set of
outgoing objects. Up to subleading order in the expansion in  powers of the soft graviton frequency,
the result depends  solely on the trajectories and the spin angular momenta
of incoming and outgoing objects (which may also include radiation), without
requiring any detailed knowledge of how the objects travelled during the scattering process 
or what kind of
forces acted between the objects during the scattering. 
In space-time dimensions larger than four,
this was proved in \cite{1801.07719} by taking the classical limit of the quantum multiple 
soft graviton theorem\cite{weinberg2,1103.2981,1401.7026,1404.4091,1405.3533,1406.6987,
1408.2228,1706.00759,
1503.04816,1504.05558,1607.02700,1707.06803,1809.01675,1811.01804}.
Our goal in this
paper will be to 
show that classical soft graviton theorem can be derived directly using the
equations of motion of general relativity coupled to 
matter. Our results will be valid in any general coordinate invariant theory of gravity
coupled to interacting matter, possibly including higher derivative terms in the action.

The strategy we follow will be to take all but the linearized terms in the Einstein's equation to the right hand side
and regard the right hand side as the total energy momentum tensor of the system. 
This includes the energy momentum tensor of the gravitational field as defined in \cite{weinberg_book}.
We can then `solve' the 
equations by taking the convolution of the flat space retarded Green's 
function with the energy momentum tensor. The domain of integration 
is then divided into two parts: a large but finite spatial volume around the region where the
scattering takes place, and the region outside this volume. In the outer region we approximate the energy
momentum tensor by that of free particles corresponding to the asymptotic 
incoming and outgoing particles and radiation. In the inner
region the energy momentum tensor, including the non-linear terms in the Einstein's equation, are complicated,
but we determine the low frequency gravitational radiation
from this region simply by using local conservation laws. 
As we show below, the sum of the two contributions is independent of the precise division of the
space-time regions we choose, and is given solely by the asymptotic 
trajectories and spin angular momenta of the incoming and the outgoing particles.

We now summarize our main results. In the following we shall refer to the incoming and outgoing
objects
involved in the scattering as particles, even though we do not assume that they are
structureless objects -- even black holes, stars and bound binary systems
will be counted as particles.
This is justified by the fact that while describing the coupling of a gravitational field of wavelength 
much larger then the characteristic size of the objects, we can approximate the stress tensor 
for any finite size gravitating object by the stress tensor of a point particle with (generically) infinitely many 
multipole moments.
We denote the asymptotic 
trajectory of the $a$-th particle by
\be
r_{(a)} = c_{(a)} +  V_{(a)}\, \sigma_a\, ,
\ee
where $c_{(a)}$ and $V_{(a)}$ are constant $D$-dimensional vectors
and $\sigma_a$ is an appropriately normalized affine parameter.
We also denote by 
$p_{(a)}\propto V_{(a)}$ the momentum of the $a$-th particle, and by 
$\Sigma_{(a)\alpha\beta}$ the spin angular momentum carried by the $a$-th particle,
both counted with + sign if ingoing and $-$ sign if outgoing. Operationally, 
$c_{(a)}$, $V_{(a)}$, $p_{(a)}$ and
$\Sigma_{(a)}$ can be defined through the energy momentum tensor 
carried by the particle
as given in \refb{etexp}.
If $h_{\alpha\beta}=(g_{\alpha\beta}-\eta_{\alpha\beta})/2$ denotes metric fluctuation, then we
define\footnote{Even though we express the metric as $\eta_{\mu\nu}+2\, h_{\mu\nu}$,
we do not assume that $h_{\mu\nu}$ is small, except in the asymptotic region.}
\be
e_{\alpha\beta}(x) \equiv h_{\alpha\beta}(x) - {1\over 2} \eta_{\alpha\beta} \, \eta^{\mu\nu} \, h_{\mu\nu}(x)\, ,
\ee
\be
\tilde e_{\alpha\beta}(\omega,\vec x) \equiv \int dx^0\, e^{i\omega x^0} \, e_{\alpha\beta}(x)
= e^{i \, \omega\, |\vec x|} \int du\, e^{i\omega u} \, e_{\alpha\beta}(x), \qquad u\equiv x^0-|\vec x|\, ,
\ee
\be
k\equiv -\omega \left(1, {\vec x\over |\vec x|}\right), \quad \NN \equiv
{1\over 2}\, \left({1\over 2\pi i |\vec x|}\right)^{(D-2)/2} \, \omega^{(D-4)/2} \, .
\ee
We show that for large $|\vec x|$, the small $\omega$ expansion of
$\tilde e_{\alpha\beta}$ is given by (up to gauge transformations):
\ben\label{esoftfinint}
\tilde e_{\alpha\beta}(\omega, \vec x)  &=& 
\NN\, e^{i \, \omega\, |\vec x|} \,  \left[ \sum_a  {p_{(a)\alpha} p_{(a)\beta}\over
k. p_{(a)}} - i\, \sum_a {1\over k.p_{(a)}}  p_{(a)(\alpha} k^\gamma
J_{(a)\beta)\gamma} +\OO(\omega)
\right] \nonumber \\ && \hskip .5in
+\OO\left( {1\over |\vec x|^{D/2}}, {1\over |\vec x|^{(D-3)}}\right)\, ,
\een
where
\be\label{edefJint}
J_{(a)\gamma\alpha}=\left\{ c_{(a)\gamma} p_{(a)\alpha} - c_{(a)\alpha} p_{(a)\gamma}
+\Sigma_{(a)\gamma\alpha}\right\}\, ,
\ee
denotes the total angular momentum carried by the $a$-th particle, with the first two terms giving the orbital
contribution and the last term giving the spin
contribution. In \refb{esoftfinint} we have set $8\pi G=1$. 
Here, and in the rest of the paper, 
all indices are raised and lowered by the Minkowski metric and all 
scalar products are also
defined using the Minkowski metric.
As indicated in the last line of \refb{esoftfinint}, the order of the error is larger of $|\vec x|^{-D/2}$
and $|\vec x|^{-(D-3)}$. An important feature of \refb{esoftfinint} is that the result does not depend on
any details of the actual scattering process or the nature of the interactions involved during the scattering.
The leading term in \refb{esoftfinint}, associated with the first term inside the
square bracket, agrees with the results obtained in 
\cite{1702.00095,1712.00873}. 

If a significant amount of momentum and / or angular momentum is carried away by the outgoing scalar, electromagnetic and / or gravitational
radiation, then the sum over $a$ in \refb{esoftfinint} also includes the contribution due to radiation.
An explicit form of this contribution may be written as:
\be\label{esoftfieldint}
-\NN\, e^{i \, \omega\, |\vec x|} \,  \left[ \int d\hat n' \,  {A_\alpha A_\beta\over k.A} -i  \int d\hat n' \, {1\over k.A} \, k^\gamma\, A_{(\alpha} B_{\beta)\gamma}
\right]\, ,
\ee
where $\int d\hat n'$ denotes angular integration, and
\ben\label{eabexpint}
&& A^\alpha(\hat n') =  \lim_{r'\to\infty} 
r^{\prime D-2}\int dt' \, \hat n'_i \, T_R^{i\alpha}(x'), \nonumber \\
&& B^{\alpha\beta}(\hat n') =   \lim_{r'\to\infty} 
r^{\prime D-2} \int dt'\, \hat n'_i\, \left\{ x^{\prime\alpha} T_R^{i\beta}(x')- x^{\prime\beta} T_R^{i\alpha}(x')\right\}, 
\nonumber \\
&& \hskip .2in r' \equiv |\vec x^{\,\prime}|, \quad
\hat n' \equiv {\vec x^{\,\prime}\over |\vec x^{\,\prime}|}\, .
\een
$T_{R\mu\nu}$ is the contribution to the symmetric energy-momentum tensor due to
massless fields.
Note that there is no ambiguity regarding the definition of $T_{R\mu\nu}$ for the
gravitational field -- it is what we get by taking the non-linear terms in the Einstein's equation to the right
hand side\cite{weinberg_book}. Explicit form of $T_{R\mu\nu}$ for massless scalar, vector and
gravitational field to the required order has been given in appendix \ref{sb}.
Physically $A^\alpha(\hat n')$ and $B^{\alpha\beta}(\hat n')$ represent 
respectively the total flux of outgoing momentum and
angular momentum\cite{1203.0452} of radiation along the direction $\hat n'$.
The overall minus sign in \refb{esoftfieldint}
reflects that in \refb{esoftfinint} the momenta and angular momenta are counted as
positive if ingoing, whereas $A^\mu(\hat n') \, d\hat n'$ and 
$B^{\mu\nu}(\hat n') \, d\hat n'$  represent outgoing flux.

With \refb{esoftfieldint} present on the right hand side of \refb{esoftfinint}, both sides of the equation
contain the gravitational field $\tilde e_{\alpha\beta}$. However one can easily check that the
contribution to \refb{esoftfieldint} from $\tilde e_{\alpha\beta}$ of frequency of order $\omega$ or less
is suppressed by higher powers of $\omega$ and therefore does not produce 
any order $\omega^0$
terms. Therefore \refb{esoftfinint} with  \refb{esoftfieldint} included can be regarded as an equation that
determines the low frequency component of the gravitational radiation in terms of its finite frequency
component and other asymptotic data. Our result is similar in spirit to the
memory effect in four dimensions (see \cite{1003.3486} for a review)
where the memory term is determined 
in terms of finite frequency gravitational radiation and other asymptotic data.

With some work,
this approach to deriving classical soft theorem may be extended to one higher order in
expansion in the soft
frequency $\omega$, by including the next order terms in the expansion \refb{etexp} of the energy
momentum tensor of matter in the far region. 
However the corresponding coefficients of expansion will not be universal, -- they will
depend on the detailed properties of the incoming and the outgoing objects. 
Based on the quantum results of \cite{1706.00759},
we expect the corrections to  $\tilde e_{\alpha\beta}$
at the next order to be of the form:
\be\label{eexpect}
-{1\over 2} \NN\, e^{i \, \omega\, |\vec x|} \,  \sum_a  {1\over
k. p_{(a)}} k^\gamma k^\delta \left[ J_{(a)\alpha\gamma} J_{(a)\beta\delta}
+ B_{(a)\alpha\gamma\beta\delta}\right]\, ,
\ee
where $B_{(a)\alpha\gamma\beta\delta}$ is some tensor that depends on the
properties of the $a$-th external state, but does not depend on the details of the
scattering process.\footnote{Results of \cite{1709.06016,1812.06895} in four 
dimensions suggest that the coefficient $B_{(a)\alpha\gamma\beta\delta}$ vanishes if the $a$-th external state
represents a rotating black hole.}
$B_{(a)\alpha\gamma\beta\delta}$ is anii-symmetric under $\alpha\leftrightarrow\gamma$ and also under
$\beta\leftrightarrow\delta$, and symmetric under $(\alpha\gamma)\leftrightarrow (\beta\delta)$.
This approach will break down at the next order due to the
ambiguity described in \refb{eobstr} in determining the contribution to $\tilde e_{\alpha\beta}$ from the near region.
This is in agreement with the corresponding results in quantum soft 
graviton theorem described in \cite{1706.00759}.

A generalization of \refb{esoftfinint}, including the contribution from massless particles,
exists in four space-time dimensions as well\cite{1804.09193,1808.03288},  
but due to the existence of long range electromagnetic and gravitational 
forces, the actual formula takes a different form
(see eq.(2.2) and (2.6) 
of \cite{1808.03288} for the general 
formula). In particular the subleading terms 
now have contribution proportional to $\ln\omega$. As in the case of $D>4$, the
four dimensional formula has been obtained by taking the classical limit of quantum
soft graviton theorem. 
This has also been verified in explicit examples where independent computation of soft
gravitational radiation during classical scattering has been performed\cite{peters,1812.08137,1901.10986}.
We expect that a direct classical derivation of this formula
should be possible along the lines discussed in the paper, but the analysis will
have to be more complicated due to the reasons described below
eq.\refb{e1fin}.

The rest of the paper is organized as follows. In \S\ref{s1} we prove the classical soft theorem
\refb{esoftfinint}, assuming that the contribution due to the radiation can be treated as a flux of massless
particles. In \S\ref{sfield} we derive the radiation contribution \refb{esoftfieldint} explicitly by analyzing the 
soft radiation sourced by the energy momentum tensor of massless fields. The two appendices
provide some technical results on the asymptotic growth of massless fields that is used in computing the
contribution to $\tilde e_{\alpha\beta}$ due to radiation.

\sectiono{Classical soft theorem} \label{s1}

We consider the situation in which a set of objects enter a given region $\RR$ in space,
interact among themselves, and then disperse.  Our goal will be to compute the spectrum of low frequency
gravitational waves emitted during this process.
Decomposing the metric $g_{\mu\nu}$ as $\eta_{\mu\nu}+2\, h_{\mu\nu}$, 
we express the
Einstein's equation as
\be \label{eeom}
\p_\rho \p^\rho h_{\mu\nu} - \p^\rho \p_\nu h_{\mu\rho} - \p^\sigma \p_\mu h_{\sigma\nu}
+ \p_\mu \p_\nu h_{\rho}^{~\rho} = -\left\{ T_{\mu\nu} - {2\over D-2} \eta_{\mu\nu} \, T_\rho^{~\rho}\right\}\, ,
\ee
where we have set $8\pi G=1$.
$T_{\mu\nu}$ contains
contribution from the matter energy momentum tensor as well as all the non-linear terms in the
Einstein's equation. Bianchi identity ensures that 
$T_{\alpha\beta}$ satisfies the conservation law (see {\it e.g.} \cite{weinberg_book}):
\be\label{econs}
\p^\alpha T_{\alpha\beta}=0\, .
\ee
Choosing de Donder gauge, 
\be
\p^\rho h_{\rho\nu} - {1\over 2} \p_\nu h_\rho^{~\rho}=0\, ,
\ee
and defining $e_{\mu\nu}$ through the equation
\be
e_{\mu\nu} = h_{\mu\nu} - {1\over 2} \eta_{\mu\nu} \, h_\rho^{~\rho}\, ,
\ee
a `solution' to \refb{eeom} may be written as
\be\label{esol}
e_{\alpha\beta}(x)  = 
-\int d^D x' \, G_r(x, x')\, T_{\alpha\beta}(x')\, ,
\ee
where $G_r$ denotes the flat space retarded Green's function 
\be \label{egrr}
G_r(x, x') = \int {d^D\ell\over (2\pi)^D} \, e^{i\ell.(x-x')} \, {1\over (\ell^0+i\eps)^2 - \vec \ell^2}\, .
\ee
We should note however that \refb{esol} should be regarded as an identity 
involving $e_{\alpha\beta}$ instead of a solution, since the right hand side of the equation also
involves $e_{\mu\nu}$ through $T_{\alpha\beta}$.

If we define
\be
\tilde e_{\alpha\beta} (\omega, \vec x) \equiv \int dx^0 \, e^{i\omega x^0}\, 
e_{\alpha\beta}(x^0, \vec x) 
\, ,
\ee
then using \refb{esol}, \refb{egrr} we get
\be
\tilde e_{\alpha\beta} (\omega, \vec x) 
=-\int d^D x' \int {d^{D-1}\vec \ell\over (2\pi)^{D-1}} \, e^{i\omega x^{\prime 0} + i\vec \ell. (\vec x-\vec x^{\,\prime})}\,
{1\over (\omega+i\eps)^2 -\vec\ell^{\, 2}}\, T_{\alpha\beta}(x')\, .
\ee
We now decompose $\vec\ell$ into its component $\ell_\parallel$ 
along $\vec x-\vec x'$ and $\vec\ell_\perp$ transverse to $\vec x-\vec x'$.
For large $|\vec x|$, we can evaluate the integration over $\ell_\parallel$ 
by closing the integration contour in the upper half plane and picking up the
residue from the pole at 
$\ell_\parallel = \sqrt{(\omega+i\eps)^2 - \vec\ell_\perp^2}$. 
After this the integration over $\vec\ell_\perp$ 
can be done using saddle point method, with the saddle
point occurring at $\vec\ell_\perp=0$. The result takes the simple 
form\cite{1801.07719}:
\be \label{emain}
\tilde e_{\alpha\beta}(\omega, \vec x) 
\simeq i\, \NN\, e^{i \, \omega\, |\vec x|} \, \int d^D x' \, e^{ik.x'}\, T_{\alpha\beta}(x')
+\hbox{boundary terms at $\infty$}\, ,
\ee
where
\be\label{esol1}
\NN =\left({\omega\over 2\pi i |\vec x|}\right)^{(D-2)/2} \, {1\over 2\omega}, 
\qquad k = -\omega\left(1, {\vec x\over |\vec x|}\right) \, ,
\ee
and the boundary terms at infinity have to be adjusted to make this integral well-defined.
$\simeq$ denotes equality up to terms containing higher powers of $1/|\vec x|$.

Let us now suppose that we have a classical scattering process in which the interaction takes place mainly
around the origin of the spatial coordinates $\vec x^{\,\prime}$.
We shall evaluate \refb{emain} by dividing the domain of integration
over the spatial coordinates $\vec x^{\, \prime}$
into two parts: $|\vec x^{\,\prime}|> L$ and $|\vec x^{\,\prime}|\le L$, and call the corresponding contributions
$\tilde e^1_{\alpha\beta}$ and $\tilde e^2_{\alpha\beta}$ respectively.
Here $L$ is a large but finite number so that
the interaction takes place mainly in the region $|\vec x^{\,\prime}|\le L$. We do not need to assume anything
about the kind of interactions that take place in this region, except that they must be consistent with
the conservation laws.
Outside this region we only need to take into account the effect of long range gravitational and
electromagnetic fields, and even these can be treated perturbatively.
We shall assume that all initial
and final particles move with finite non-zero velocity so that in the far past and far future the region
$|\vec x^{\,\prime}|\le L$ is nearly empty. This can always be achieved by choosing an appropriate Lorentz
frame. However, since our final formula will be written in a Lorentz covariant form, it will also be valid in the frame
where some of the initial or final state particles are at rest.

In the region $|\vec x^{\,\prime}|\le L$, we do not make any 
assumption about $T_{\mu\nu}$ except its conservation laws. We shall see that this is
sufficient to extract the relevant contribution to the integral from this region.
On the other hand in the
interval $|\vec x^{\,\prime}|> L$, $T_{\mu\nu}$
will be taken to be the energy momentum tensor  of free particles
whose quantum numbers are the same as those of the incoming and the outgoing 
particles. We do not however assume that the particles are structureless:
the effect of the internal structure of the particle is encoded in the fact that the 
energy momentum tensor of the particles is allowed to have derivatives of delta functions
localized on the trajectory besides the leading term proportional to the 
delta function\cite{Tulczyjew,Dixon,0511061,0511133,0604099,0804.0260,1709.06016,
1812.06895}.
In this case the
Fourier
transform $\wt T_{\mu\nu}(k)$ 
of the energy momentum tensor will have an expansion in powers of $k$, with the
expansion coefficients encoding the internal structure of the particle. 
As long as we consider wavelengths large compared to the sizes of the particles, this
expansion will be valid even for big objects like neutron stars, black holes or binary
systems. 
For our analysis we
shall only need the first two coefficients in the expansion, 
which are determined in terms of the momentum and
spin of the particle. The precise expression will be given shortly.

\begin{figure}
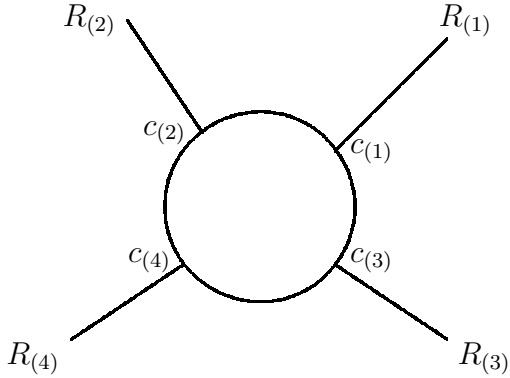


\begin{center}

\figonea

\end{center}

\caption{Geometry of a scattering process.  \label{fig1}}

\end{figure}

We begin our analysis in the $|\vec x^{\,\prime}|> L$ region. In this region
we have, to first order in the expansion in derivatives of delta 
function,\footnote{For the trajectory $r_{(a)}(\sigma_a)= V_{(a)}\sigma_a+ c_{(a)}$, 
the right hand side of \refb{etexp} 
can be shown to be invariant under the transformation
$c_{(a)}\to c_{(a)}+b_{(a)}$, $\Sigma_{(a)\alpha\beta}\to \Sigma_{(a)\alpha\beta}-b_{(a)\alpha} p_{(a)\beta}
+ p_{(a)\alpha} b_{(a)\beta}$, for any constant vector $b_{(a)}$. Therefore neither $c_{(a)}$ nor
$\Sigma_{(a)\alpha\beta}$ are unambiguously defined. 
We can fix this ambiguity by imposing some conditions on 
$\Sigma_{(a)\alpha\beta}$, {\it e.g.} $p_{(a)}^\alpha \Sigma_{(a)\alpha\beta}=0$.
In any case, $J_{(a)\alpha\beta}$ defined in
\refb{edefJint} is unambiguous. }
(see {\it e.g.} \cite{1712.09250}):
\be \label{etexp}
T_{\alpha\beta}(x') = \sum_a \int d\sigma_a\, 
\left[V_{(a)\alpha} p_{(a)\beta} \delta^{(D)} (x' - r_{(a)}(\sigma_a)) 
 + V_{(a)(\alpha} \Sigma_{(a)\beta)\gamma}
\p^{\prime\gamma}\delta^{(D)} (x' - r_{(a)}(\sigma_a)) \right]\, ,
\ee
where the sum over $a$ runs over all the incoming and outgoing particles, 
$r_{(a)}(\sigma_a)$
denotes the trajectory of the $a$'th particle with $\sigma_a$ labelling an
appropriately normalized affine parameter along
the trajectory up to a sign,
$p_{(a)}$
and $\Sigma_{(a)}$ 
are respectively the $D$-momentum and the spin angular momentum
of the $a$'th particle, both counted with
positive sign for ingoing particles and negative sign for outgoing particles, and
$V_{(a)}=dr_{(a)}/d\sigma_a$. 
Since in our notation the 
ingoing momenta are positive, we take $\sigma_a$ to increase from $-\tau_a$ 
at the outer end
to $0$ on the surface $|\vec x^{\, \prime}|= L$.
In this notation $V_{(a)} = K_{(a)} p_{(a)}$ for some positive constant 
$K_{(a)}$ and the trajectory begins 
at some cut-off point $R_{(a)}\equiv
r_{(a)}(-\tau_a)$ at the outer end and ends at a point
$c_{(a)}\equiv r_{(a)}(0)$ on the surface $|\vec x^{\,\prime}|= L$. 
This has been illustrated
in Fig.~\ref{fig1}. Therefore by definition 
$|\vec c_{(a)}|=L$.
$(\alpha\ldots\beta)$ denotes symmetrization, with the convention:
\be
A_{(\alpha} B_{\beta)} =  {1\over 2} \left(A_\alpha B_\beta + B_\alpha A_\beta\right)\, .
\ee
In the sum over $a$ in \refb{etexp}, 
we include the effect of outgoing radiation (gravitational or electromagnetic)
by regarding them as a flux of massless particles.\footnote{This will be justified in detail in \S\ref{sfield}.} 
Therefore the sum over $a$ includes 
an angular integration over outgoing finite frequency radiation.
In the special case where the total
energy carried away by radiation is small compared to the energies of the massive
objects involved in the scattering, the contribution due to radiation in the sum in \refb{etexp}
can be ignored.

There are of course higher order terms in the expansion \refb{etexp} containing more derivatives of 
$\delta^{(D)} (x' - r_{(a)}(\sigma_a))$, but they will not contribute to the soft theorem to subleading order.
This can be seen as follows. First we see that when we substitute \refb{etexp} into \refb{emain}, the
$x'$ integral in the first term gets localized at $r_{(a)}(\sigma_a)$, but the $\sigma_a$ integration produces a linearly divergent
term from the large $\sigma_a$ region in the $k\to 0$ limit. This divergence is regulated by the
$e^{ik.x'}$ term in \refb{emain}, producing an inverse power of $k$. This gives the leading term. Since
the second term in \refb{etexp}
contains a derivative of $\delta^{(D)} (x' - r_{(a)}(\sigma_a))$, we can first integrate by parts,
bringing down a factor of $k.V_{(a)}(\sigma_a)$ and then repeat the previous argument. As a result this
term is of order unity and begins contributing at the subleading order. Terms involving higher derivatives
of $\delta^{(D)} (x' - r_{(a)}(\sigma_a))$ will bring down more powers of $k$. Therefore they will not contribute at
the subleading order.

Using \refb{etexp}, the contribution $\tilde e^1_{\alpha\beta}$ is given by:
\ben\label{ebou}
\tilde e^1_{\alpha\beta}(k) &\equiv& i\, \NN\, e^{i \, \omega\, |\vec x|} \, \int_{|\vec x^{\,\prime}|> L} 
d^D x' \, e^{ik.x'}\, T_{\alpha\beta}(x') +\hbox{boundary terms at $\infty$}\nonumber\\ &=&
i\, \NN \,  e^{i\omega |\vec x|} \, 
\sum_a \int_{-\tau_a}^0 d\sigma_a e^{ik.r_{(a)}(\sigma_a)}
\left\{V_{(a)\alpha} p_{(a)\beta} -i \, k^\gamma \,
V_{(a)(\alpha}\, \Sigma_{(a)\beta)\gamma}\right\}
\nonumber\\  &&
+\, i\, \NN \,  e^{i\omega |\vec x|} \sum_a \,  e^{ik.c_{(a)}}\, {1\over \tilde n_{(a)}.V_{(a)}}\, 
V_{(a)(\alpha}\, \Sigma_{(a)\beta)\gamma}\, \tilde n_{(a)}^\gamma
\nonumber \\ &&
+ \, \hbox{boundary terms at $\infty$}
\, ,
\een
where we define,
\be 
\tilde n' = (0, \vec x^{\,\prime} / |\vec x^{\,\prime}|), \quad 
\tilde n_{(a)} = (0, \vec c_{(a)}/ |\vec c_{(a)}|)
\, .
\ee
The term in the penultimate line of \refb{ebou} is a boundary term at $|\vec x^{\,\prime}|=L$ 
that arises from having to integrate by
parts the term involving $\p'_\gamma$ in \refb{etexp}. This can be seen as follows. 
First we represent the boundary term as
\be
- i\, \NN \,  e^{i\omega |\vec x|} \sum_a \,  e^{ik.c_{(a)}}\, \int d^D x' \,  \int d\sigma_a \,
\delta(|\vec x^{\,\prime}|-L)\, 
 V_{(a)(\alpha}\, \Sigma_{(a)\beta)\rho}\, \tilde n^{\prime\rho} \, \delta^{(D)} (x' - r_{(a)}(\sigma_a)) \, .
 \ee
We now carry out the
integration over $x'$ using the $\delta^{(D)} (x' - r_{(a)}(\sigma_a))$ factor. This replaces 
$\delta(|\vec x^{\,\prime}|
-L)$ by $\delta(|\vec r_{(a)}(\sigma_a)|-L)$. We then carry out the $\sigma_a$ integration using this
delta function, which has support at $\sigma_a=0$ since $|\vec r_{(a)}(0)|=|\vec c_{(a)}|
=L$. This generates a factor of $-{1/ \tilde n_{(a)}.V_{(a)}}$, with the minus sign reflecting the fact that
$V_{(a)}$ is counted as positive if ingoing. 

There are similar terms from the
outer boundary where $r_{(a)}(\sigma_a)$ takes value $R_{(a)}$, but these
terms are absorbed into the boundary terms at $\infty$ in the last line 
of \refb{ebou}.
Using the trajectory equation
$r_{(a)}(\sigma_a) = c_{(a)} + V_{(a)} \, \sigma_a$, 
we can carry out the integration over $\sigma_a$ in \refb{ebou}. If we now use
the relation 
$V_{(a)} = K_{(a)}\, p_{(a)}$ for some positive constant $K_{(a)}$, we can express
\refb{ebou} as:
\be \label{e1fin}
\tilde e^1_{\alpha\beta}(k) = \NN \,  e^{i\omega |\vec x|} \, 
\sum_a e^{i k. c_{(a)}} \left[ {1\over
k. p_{(a)}} \left\{ p_{(a)\alpha} p_{(a)\beta} - i\, k^\gamma\, 
p_{(a)(\alpha}\, \Sigma_{(a)\beta)\gamma}
\right\}  + i\, {1\over \tilde n_{(a)}.p_{(a)}}\, 
p_{(a)(\alpha}\, \Sigma_{(a)\beta)\gamma}\, \tilde n_{(a)}^\gamma\right]\, .
\ee
In arriving at this expression we have cancelled all terms proportional to 
$e^{ik.R_{(a)}}$ by appropriate choice of the boundary terms at $R_{(a)}$.

Let us estimate the error we made in the above calculation by taking $T_{\mu\nu}$ to be the
energy momentum tensor produced by free particles. Since we are computing the result to subleading order
$\omega^0$, we shall estimate the error to this order.  
First error stems from the fact that the
particles are not free, but are under the influence of each other's long range gravitational (and
possibly electromagnetic) fields. These forces fall off as $1/r^{D-2}$ 
when the distance between the
particles is of order $r$ -- with all distances being measured with respect to the flat metric. 
Integrating this once we see that the correction to $P_\alpha$ (and
hence also $V_\alpha$) fall off as $1/r^{D-3}$. Therefore the integral of the error over part of
the trajectory from $r$ to $\infty$ will fall off as $1/r^{D-4}$. Since $r\ge L$ in the integration region for
evaluating $\tilde e^1_{\alpha\beta}$, the
net error in the  computation of $\tilde e^1_{\alpha\beta}$ 
is
bounded by $1/L^{D-4}$. For $D>4$, this error vanishes in the large $L$ limit. 

There may also be contributions to $\tilde e^1_{\alpha\beta}$ from
$T_{\alpha\beta}$ stored in the long range fields (gravitational and electromagnetic). 
We shall show in  \S\ref{sfield} that this contribution can be included
in the sum over $a$ in \refb{e1fin} by regarding the radiative field
contribution as a sum over the flux of massless particles. 
In the final expression, 
the additional contribution to $\tilde e_{\alpha\beta}$
due to radiation will be given by \refb{esoftfieldint}.

We now turn to the contribution $\tilde e^2_{\alpha\beta}$ to \refb{emain} from the 
$|\vec x^{\, \prime}|\le L$ region. We have:
\be \label{eintan}
\tilde e^2_{\alpha\beta}(k) =  i\, \NN\, e^{i \, \omega\, |\vec x|} \, 
\int_{|\vec x^{\, \prime}|\le L} d^D x' \, e^{ik.x'}\, T_{\alpha\beta}(x')\, .
\ee
This gives:
\be \label{e2.18}
i\, k^\alpha \tilde e^2_{\alpha\beta}(k) =  i\, \NN\, e^{i \, \omega\, |\vec x|} \, 
\int_{|\vec x^{\, \prime}|\le L} d^D x' \, 
{\p\over \p x^{\prime \alpha}} \left(e^{ik.x'}\right)\, T^{\alpha}_{~\beta}(x')\, .
\ee
Using integration by parts and the conservation law \refb{econs}, this can be expressed as
\ben \label{econd1.1}
&& i\, k^\alpha \tilde e^2_{\alpha\beta}(k) = i\, 
\NN\, e^{i \, \omega\, |\vec x|} \, \int_{|\vec x^{\,\prime}|=L} d^{D-1} x' \, e^{ik.x'}\, T_{\gamma\beta}(x')\,
\tilde n^{\prime\gamma}, \nonumber \\ && \hskip 1in
  \int_{|\vec x^{\,\prime}|=L} d^{D-1} x' \, \cdots
\equiv \int d^D x' \delta(|\vec x^{\,\prime}|-L)\, \cdots
\, .
\een
We can evaluate the right hand side by
noting that on the boundaries $|\vec x^{\,\prime}|= L$ the energy momentum
tensor may be approximated by those of the free particles entering and leaving
the region $|\vec x^{\,\prime}|= L$. 
We now use \refb{etexp} to express \refb{econd1.1} as,
\ben\label{econd}
i\, k^\alpha \tilde e^2_{\alpha\beta}(k) &=&  i\, 
\NN\, e^{i \, \omega\, |\vec x|} \, \int_{|\vec x^{\,\prime}|=L} d^{D-1} x' \, e^{ik.x'}\, 
\sum_a  \int d\sigma_a\, 
\tilde n^{\prime\gamma} \, V_{(a)\gamma}(\sigma_a) p_{(a)\beta}(\sigma_a) 
\delta^{(D)} (x' - r_{(a)}(\sigma_a)) 
\nonumber \\ && \hskip -1.1in
 + {i\over 2}\, 
\NN\, e^{i \, \omega\, |\vec x|} \, \Bigg[\int_{|\vec x^{\,\prime}|=L} d^{D-1} x' \, e^{ik.x'}\, 
\sum_a \int d\sigma_a\, \tilde n^{\prime\rho}\, 
\left\{ V_{(a)\rho}(\sigma_a) \Sigma_{(a)\beta\gamma}(\sigma_a) +  V_{(a)\beta}(\sigma_a)
 \Sigma_{(a)\rho\gamma}(\sigma_a)\right\} \nonumber \\ &&
\hskip 1in \p^{\prime\gamma}\delta^{(D)} (x' - r_{(a)}(\sigma_a)) 
\Bigg]\, . 
\een
We manipulate this expression using the following
steps.
\begin{enumerate}
\item Replace $\int_{|\vec x^{\, \prime}|=L} d^{D-1} x' \cdots$ by $\int d^{D} x'
 \delta(|\vec x^{\,\prime}|-L)\, \cdots$.
\item Integrate the $\p^{\prime\gamma}$ term by parts. This acts on $e^{ik.x'}$ and brings
down a factor of $-ik^\gamma$. It also acts on the
$ \delta(|\vec x^{\,\prime}|-L)$ and gives
$-\tilde n^{\prime\gamma}  \delta'(|\vec x^{\,\prime}|-L)$.
\item Once the $\p^{\prime\gamma}$ factor has been removed from
$\delta^{(D)} (x' - r_{(a)}(\sigma_a))$, we can use this delta function to perform the 
$x'$ integration. This sets $x' = r_{(a)}(\sigma_a)$ in all the expressions, including in
the $\tilde n^{\prime\gamma}  \delta'(|\vec x^{\,\prime}|-L)$ term and in the
$e^{ik.x'}$ term.
\item We now use the fact that the solution to relations $|\vec r_{(a)}(\sigma_a)|=L$ 
is $\sigma_a=0$ to write
\be
\delta(|\vec r_{(a)}(\sigma_a)| - L) = - \, (\tilde n_{(a)}. V_{(a)})^{-1} \delta(\sigma_a),
\quad \delta'(|\vec r_{(a)}(\sigma_a)| - L) = - (\tilde n_{(a)}.V_{(a)})^{-2}
 \delta'(\sigma_a)\, .
\ee
The minus sign reflects the fact that
$V_{(a)}$ is counted as positive if ingoing. 
\item We can now perform the integration over $\sigma_a$ using these delta functions.
In particular the $\delta'(\sigma_a)$ term will generate a $\sigma_a$ derivative of
$-e^{ik.r_{(a)}(\sigma_a)}$, producing a term proportional to $-ik.V_{(a)}$.
\item  We now replace all factors of $V_{(a)}$ by $K_{(a)} \, p_{(a)}$ with $K_{(a)}$
being a positive constant.
\end{enumerate}
The net outcome of these manipulations is the relation:
\ben \label{econd2pre}
i\, k^\alpha \tilde e^2_{\alpha\beta}(k) &=& -i\, \NN\, e^{i \, \omega\, |\vec x|} \, 
\sum_a e^{i k. c_{(a)}}\, \bigg\{ 
p_{(a)\beta} - {i\over 2} k^\gamma \Sigma_{(a)\beta\gamma}
+{i\over 2} k^\gamma  {p_{(a)\beta}\over \tilde n_{(a)} .p_{(a)}}
\Sigma_{(a)\gamma \rho}\, \tilde n_{(a)}^{\rho} 
\nonumber \\ && \hskip 2in +{i\over 2} {k.p_{(a)}\over  \tilde n_{(a)} .p_{(a)}}\, 
\Sigma_{(a)\beta \rho}
\, \tilde n_{(a)}^{\rho}
\bigg\}\, .
\een
Expanding the $e^{i k. c_{(a)}}$ factor in powers of $k.c_{(a)}$ and
using the momentum conservation law 
\be\label{emattcons}
\sum_a p_{(a)}=0\, ,
\ee 
we can express \refb{econd2pre} as
\ben \label{econd2}
i\, k^\alpha \tilde e^2_{\alpha\beta}(k) &=& -i\, \NN\, e^{i \, \omega\, |\vec x|} \, 
\sum_a \, \Bigg( i \, k.c_{(a)}\, p_{(a)\beta} - {i\over 2} k^\gamma \Sigma_{(a)\beta\gamma}
+{i\over 2} k^\gamma  {p_{(a)\beta}\over \tilde n_{(a)} .p_{(a)}}
\Sigma_{(a)\gamma \rho}\, \tilde n_{(a)}^{\rho} \nonumber \\ &&
\hskip 1.5in +{i\over 2} {k.p_{(a)}\over  \tilde n_{(a)} .p_{(a)}}\, 
\Sigma_{(a)\beta \rho}
\, \tilde n_{(a)}^{\rho}
\Bigg)+\OO(\omega^2)\, .
\een

Since in the definition \refb{eintan} of $\tilde e^2_{\alpha\beta}$ the integration over 
$\vec x^{\,\prime}$ 
is confined to a finite region (which also effectively 
makes the integration over $x^{\prime\, 0}$ bounded since by assumption the region $|\vec x^{\,\prime}|=L$
becomes empty for large $|x^{\prime 0}|$), 
$\tilde e^2_{\alpha\beta}$ is an analytic function of $k^\mu$ near $k=0$ and should admit a Taylor series expansion
in $k^\mu$.
We propose the following solution for $\tilde e^2_{\alpha\beta}(k)$:
\ben\label{esolp}
\tilde e^2_{\alpha\beta} &=& -i\, \NN\, e^{i \, \omega\, |\vec x|} \, 
\sum_a \left(c_{(a)\alpha} p_{(a)\beta} - {1\over 2} \Sigma_{(a)\beta\alpha}
+ {1\over 2} {p_{(a)\beta}\over  \tilde n_{(a)} .p_{(a)}}
\Sigma_{(a)\alpha \rho}\, \tilde n_{(a)}^{\rho} +{1\over 2} {p_{(a)\alpha}\over  \tilde n_{(a)} .p_{(a)}}\, 
\Sigma_{(a)\beta \rho}\, \tilde n_{(a)}^{\rho}
\right)\nonumber \\
&& \hskip 2in  + \OO(k)\, .
\een
It satisfies \refb{econd2} up to terms of order $\omega$. We also need to check that this
is symmetric under exchange of $\alpha$ and $\beta$. For this we note that angular
momentum conservation implies:
\be\label{einr1}
\sum_a \left(c_{(a)\alpha} p_{(a)\beta} - c_{(a)\beta} p_{(a)\alpha} + 
\Sigma_{(a)\alpha\beta}\right)=0\, .
\ee
Adding \refb{einr1} multiplied by $i\, \NN\, e^{i \, \omega\, |\vec x|}/2$ to \refb{esolp},
we get:
\be\label{e2fin}
\tilde e^2_{\alpha\beta} = -i\, \NN\, e^{i \, \omega\, |\vec x|} \, 
\sum_a \left\{c_{(a)(\alpha} p_{(a)\beta)} 
+ {1\over 2} {p_{(a)\beta}\over \tilde n_{(a)} .p_{(a)}}
\Sigma_{(a)\alpha\rho}\, \tilde n_{(a)}^{\rho} +{1\over 2} {p_{(a)\alpha}\over \tilde n_{(a)} .p_{(a)}}\, 
\Sigma_{(a)\beta \rho}
\, \tilde n_{(a)}^\rho
\right\}
+ \OO(k)\, ,
\ee
which is manifestly symmetric under $\alpha\leftrightarrow\beta$.

We can also argue that the solution
\refb{esolp} is unique. To see this we assume the contrary, that
there is another solution. Then the difference $d_{\alpha\beta}$ 
between the two solutions will be analytic function of $k^\mu$ near $k=0$ and will
satisfy the constraint $k^\alpha d_{\alpha\beta}=0$. It is easy to check that there is
no function $d_{\alpha\beta}(k)$ that satisfies this requirement,
is analytic at $k=0$ and does not vanish at $k=0$. The first term in the
power series expansion in $k_\mu$ that satisfies this constraint is proportional to 
\be\label{eobstr}
k^2\eta_{\alpha\beta}
-k_\alpha k_\beta\, .
\ee

Adding \refb{e1fin} and \refb{e2fin}, and expanding in powers of $k$, we get
\be\label{esoftfin}
\tilde e_{\alpha\beta} = \tilde e^1_{\alpha\beta} + \tilde e^2_{\alpha\beta} = 
\NN\, e^{i \, \omega\, |\vec x|} \,  \left[ \sum_a  {p_{(a)\alpha} p_{(a)\beta}\over
k. p_{(a)}} - i\, \sum_a {1\over p_{(a)}.k}  p_{(a)(\alpha} k^\gamma
J_{(a)\beta)\gamma}
\right] + \OO(k)\, ,
\ee
where
\be
J_{(a)\gamma\alpha}=\left\{ c_{(a)\gamma} p_{(a)\alpha} - c_{(a)\alpha} p_{(a)\gamma}
+\Sigma_{(a)\gamma\alpha}\right\}\, .
\ee
This is the classical soft graviton theorem to subleading order. We emphasis that the sum over
$a$ in \refb{esoftfin} includes integration over the flux of gravitational (and electromagnetic if any)
radiation, with $J_{(a)\gamma\alpha}$ representing the flux of angular momentum carried by the
radiation. Explicit expression for these contributions has been given in \refb{esoftfieldint}. In \S\ref{sfield}
we shall derive \refb{esoftfieldint} by directly working with massless fields instead of regarding them as
flux of massless particles.

We conclude this section by exploring the possibility of extending the analysis to higher orders 
in the frequency $\omega$ of the soft graviton:
\begin{enumerate}
\item In order to extend our computation of $\tilde e^1_{\alpha\beta}$ to higher order in $\omega$, we need to
keep terms in the expression \refb{etexp} involving higher number of derivatives of the delta function.
However the coefficients of these terms are not expected to be universal. Instead they will depend on the
internal structures of the objects involved in the scattering. Nevertheless, these contributions will still be
independent of the details of the scattering process, being sensitive only to the properties of the
incoming and the outgoing objects. For Kerr black holes in four dimensions, some of
the coefficients of higher derivatives of the delta function have been computed in
\cite{1709.06016,1812.06895}.
\item Presence of the term \refb{eobstr} in the expression for $\tilde e^2_{\alpha\beta}$ begin affecting the soft
radiation at the sub-sub-subleading order. These contributions are expected to depend on the
details of the scattering process and not just on the properties of the incoming and the 
outgoing objects. Therefore
our approach cannot unambiguously determine the low frequency gravitational
radiation in terms of the properties of the incoming and outgoing objects
beyond the sub-subleading order and further details of the theory are required to 
determine the metric.
\end{enumerate}
We have described in \refb{eexpect} the expected correction to $\tilde e_{\alpha\beta}$
at the subsubleading order in the expansion in powers of $\omega$. We hasten to add
however that this expectation is based on the quantum soft graviton theorem derived
in \cite{1706.00759}, and we have not derived \refb{eexpect} from a classical analysis.

\sectiono{Soft radiation from fields} \label{sfield}

Our goal in this section will be to compute soft radiation sourced by fields. We shall use \refb{emain} for this
computation, by dividing the integration region into the $|\vec x^{\,\prime}|> L$ and 
$|\vec x^{\,\prime}|\le L$ parts as in \S\ref{s1}. We shall divide the analysis into two parts. In the first part we
shall derive the analog of \refb{etexp} for radiation. In the second part we shall use this result to compute soft
radiation from the radiative stress tensor.

\subsection{General form of the stress tensor of radiation} \label{s2.1}

We begin by introducing some notations.
Let us first define:
\be\label{e3.1}
t'=x^{\prime 0}, \quad
r' = |\vec x^{\,\prime}|, \quad \hat n' = \vec x^{\,\prime} / r', \quad n' = (1, \hat n'), \quad \tilde n'=(0, \hat n'),
\quad u' = t'- r',
\ee 
where in \refb{e3.1}, $n'$ and $\tilde n'$ are to be regarded as contravariant vectors.
We shall denote by $\p_\mu$ the derivative $\p/\p x^{\prime\mu}$. 
We also define the transverse derivative
$\vec \p_\perp$ as follows.
If $\delta \hat n'$ denotes an infinitesimal
vector orthogonal to $\hat n'$, so that $\hat n' +\delta \hat n'$ is still a unit vector to first order,
then we define $\vec\p_\perp S$ for any function $S(\hat n')$ via the
relation:
\be \label{edefdelperps2}
S(\hat n' +\delta \hat n') = S(\hat n') + \delta \hat n' . \vec\p_\perp S, \quad \hat n' . \vec\p_\perp S=0\, .
\ee
It is easy to verify the following identities,
\be \label{e3.3s2}
\p_\mu r' = \tilde n'_\mu, \quad \p_\mu u' = -n'_\mu, \quad \p_\mu \hat n'_i =
- r^{\prime -1} (n'_i \tilde n'_\mu -\delta_{i\mu}) = r^{\prime -1} \, \eta^\perp_{\mu i},
\quad \p_{\perp i} n'_\mu = \eta^\perp_{i\mu} = \p_{\perp i} \tilde n'_\mu  \, ,
\ee
where 
\be
\eta^\perp_{\mu\nu}\equiv \eta_{\mu\nu} + n'_\mu n'_\nu - n'_\mu \tilde n'_\nu - n'_\nu \tilde n'_\mu \, ,
\ee 
is the projection operator into the space transverse to $n'$ and $\tilde n'$. This gives,
for any function $f(r', u, \hat n')$:
\be\label{eneweq}
\p_\mu f = \tilde n'_\mu \, \p_{r'} f - n'_\mu \, \p_{u'} f + {1\over r'} \eta^\perp_{\mu i} 
\p_{\perp i} f\, .
\ee

We shall now derive 
the analog of \refb{etexp} -- the asymptotic form of the stress tensor 
$T_{R\mu\nu}$ associated with massless fields,
including gravitational and electromagnetic field. As in \S\ref{s1}, the stress tensor of the
gravitational field will be defined to be whatever appears on the right hand side of the Einstein's equation
if the left hand side contains only the linear terms.
We
claim that the relevant part of the
stress tensor produced by radiation has the following expansion to order
$r^{\prime -(D-1)}$:
\ben \label{eid3s2}
T_{R\mu\nu}(x') &=& {1\over r^{\prime D-2}}\, 
R(\hat n', u')  \,  n'_\mu \, n'_\nu 
-{1\over  r^{\prime D-1}}  \,
\left\{n'_\mu R_{\perp\nu} (\hat n', u') 
+ n'_\nu R_{\perp\mu}(\hat n', u')  \right\} \nonumber \\ && 
 - {1\over  r^{\prime D-1}}  \, n'_\mu \, n'_\nu\, \vec\p_\perp.\vec R_\perp
(\hat n', u')
+ {1\over r^{\prime D-1}}\, \p_{u'} N_{\mu\nu}(\hat n', u') \, ,
 \een
 where $R_\perp(\hat n', u')$ is a vector with only transverse components:
 \be
 R_\perp = (0,\vec R_\perp), \quad \hat n'.\vec R_\perp(\hat n', u')=0\, .
 \ee 
 Furthermore, we show below that for large $|u'|$, $R$ and $R_\perp$ fall off at least as fast as 
 $|u'|^{-(D-2)}$
and
$|u'|^{-(D-3)}$ respectively.
 
 The justification for \refb{eid3s2} 
 can be given as follows. We begin with the leading term\footnote{This result
 is well known (see {\it e.g.} \cite{bondi}), but we include the argument for completeness.}
 $R\,  n'_\mu \, n'_\nu /r^{\prime (D-2)}$. Any component $\phi_a$ of a 
 massless field, irrespective of its spin,
has a leading behaviour of the form
$f_a(\hat n',u')/r^{\prime(D-2)/2}$ for large $r'$.  Furthermore, as reviewed in
appendix \ref{sa}, 
$f_a(\hat n',u')$ falls off at large $|u'|$\cite{1612.03290,1702.00095,1712.00873}.
Using \refb{eneweq} we now see that 
the leading term in $\p_\mu \phi_a$ is given by $-n'_\mu 
\p_{u'} f_a(\hat n',u')/r^{\prime(D-2)/2}$. 
Since the relevant term in the stress tensor in the
asymptotic region comes from the square of the first derivative of the field, we see that 
in order to get a term of the form $1/r^{\prime (D-2)}$ in the stress tensor we must 
take the term $n'_\mu n'_\nu \p_{u'} f_a  \p_{u'} f_b / r^{\prime(D-2)}$ and
appropriately contract the indices. Since 
$n'_\mu$ cannot contract with itself or the transverse indices, 
by taking the leading order term in the fields to carry only transverse 
polarization -- which is possible for $D>4$ -- 
we can ensure that $n'_\mu$ and $n'_\nu$ remain uncontracted and only
the transverse indices are contracted with each other. 
This shows 
that the $1/r^{\prime (D-2)}$
term in $T_{R\mu\nu}$ must be proportional
to $n'_\mu n'_\nu$, i.e.\ it takes the form given in the first term in \refb{eid3s2}. 
Using \refb{e3.3s2}, \refb{eneweq} one can show that this term satisfies the energy momentum conservation law
$\p^\mu T_{R\mu\nu}=0$ by itself.

We now turn to the subleading terms in \refb{eid3s2}. We begin by writing down the most
general expression for the order $1/r^{\prime(D-1)}$ term in $T_{R\mu\nu}$:
 \be
{1\over r^{\prime D-1}} \left[ A n'_\mu n'_\nu + (n'_\mu B_{\perp\nu} + n'_\nu B_{\perp\mu}) + 
C (n'_\mu \tilde n'_\nu
+ n'_\nu \tilde n'_\mu) + F \, \tilde n'_\mu \tilde n'_\nu + (\tilde n'_\mu G_{\perp\nu} + \tilde n'_\nu G_{\perp\mu})
+ H_{\perp\mu\nu}
\right]\, ,
\ee
where $A$, $C$ and $F$ are scalars, 
$B_{\perp\mu}$, $G_{\perp\mu}$ are transverse vectors and $H_{\perp\mu\nu}$ is a transverse symmetric tensor, 
all the quantities being functions of $\hat n'$ and $u'$. We now demand $\p^\mu T_{R\mu\nu}=0$ and use
\refb{e3.3s2}. 
We get, at order $1/r^{\prime D-1}$:
\be 
\p_{u'} C=0, \quad \p_{u'} F = 0, \quad \p_{u'} G_{\perp\nu} =0\, . 
\ee
Since for fixed $r'$ the stress tensor must vanish for $u'\to -\infty$, this gives:
\be 
C=0, \quad F=0, \quad G_{\perp\nu} = 0 \, .
\ee
Vanishing of the term proportional to $1/r^{\prime D}$ in $\p^\mu T_{R\mu\nu}$ give
\be 
A - \p_\perp.B_\perp = 0, \quad \p_{\perp\mu} H_\perp^{\mu\nu} = 0 \, .
\ee
The right hand sides of these equations could actually have terms
proportional to $u'$-derivatives of higher order coefficients of expansion, but since
such
terms can be absorbed into a redefinition of $N_{\alpha\beta}$ in \refb{eid3s2},
we ignore them. 
This brings the result almost to the desired form \refb{eid3s2} with the identification 
$B_\perp=-R_\perp$, except that we need to
show that the transverse tensor $H_{\perp\alpha\beta}$ has the form of 
the term proportional to $\p_{u'} N_{\alpha\beta}$ in \refb{eid3s2}.
This can be seen as follows:
\begin{enumerate}
\item
Since we have chosen the polarization tensors of the fields $\phi_a$ to be 
transverse (and also traceless for the radiative part of the metric), 
neither the $u'$ nor the $r'$ derivative can contract with an index of the polarization tensor.
One way to get rid of the free $u$ index from $\p_{u'}$ is to pick the other derivative to be 
$\p_{r'}$ and contract $\p_{u'}$ with $\p_{r'}$ -- we see from \refb{eneweq} that this is possible since
$n'.\tilde n'=1$. This leaves behind the indices from the polarization tensors,
which could supply the indices of the transverse tensor $H_{\mu\nu}$ (and, for the gravitational field, 
the left-over 
transverse indices can contract with each other). Since $\p_{r'}\phi_a\propto f_a(\hat n',u')\, r^{\prime -D/2}$
and $\p_{u'}\phi_a\propto \p_{u'} f_a(\hat n',u')\, r^{\prime -(D-2)/2}$, such 
contributions to $H_{\perp\mu\nu}$ will have the structure $r^{
\prime -(D-1)} \p_{u'} f_a \, f_b$ with the polarizations
appropriately contracted. It is easy to see that these terms are total derivatives in $u'$ and therefore can be
absorbed into the term proportional to $\p_{u'}N_{\alpha\beta}$ in \refb{eid3s2}. For example for gauge fields
we shall have $H_{\perp ij}\propto (f_i \p_{u'} f_j + f_j \p_{u'} f_i) = \p_{u'} (f_i f_j)$ and for the gravitational
field we have $H_{\perp ij}\propto (f_{i\ell} \p_{u'} f_{j\ell} + f_{j\ell} \p_{u'} f_{i\ell}) = \p_{u'} (f_{i\ell} f_{j\ell})$. 
\item For the gravitational field 
we need to also consider a possible contribution
to $T_{ij}$
proportional to $\p_{u'} h_{rr} \p_{u'} h_{ij}$, where $i,j$ are transverse directions
and the two $u$ indices coming from
$\p_{u'}$ are contracted with the two $r$ indices of $h_{rr}$.
Even if by a choice of gauge we take the order $r^{\prime -(D-2)/2}$ 
term in $h_{\mu\nu}$ to have only transverse components,
$h_{rr}$ could have a term of order $r^{\prime -D/2}$. Therefore, 
$\p_{u'} h_{rr} \p_{u'} h_{ij}$ could give a contribution to $T_{ij}$
of order $r^{\prime -(D-1)}$. However, as shown in eq.\refb{eb36} in appendix \ref{sb},
equations of motion forces the leading term in $h_{rr}$
to be of order
$r^{\prime -(D+2)/2}$. Therefore we cannot get a contribution of order 
$r^{\prime-(D-1)}$ to
$T_{ij}$ with $i,j$ transverse from the $\p_{u'} h_{rr} \p_{u'} h_{ij}$ term.
\end{enumerate}
This establishes \refb{eid3s2}.

In appendix \ref{sb} we shall verify \refb{eid3s2} explicitly for massless scalar, 
vector and tensor fields, where we also express \refb{eid3s2} in Bondi coordinates.
From this analysis it will also be 
clear that $R$ and $R_\perp$ contain two powers of $f_a(\hat n',u')$,
with $R$ having two $u'$ derivative acting on these factors and $R_\perp$ having one 
$u'$ derivative acting on one of the factors.  On the other hand, it follows from the analysis of 
\cite{1612.03290,1702.00095,1712.00873} 
-- rederived  in appendix \ref{sa} --
that $f_a(\hat n',u')$
falls off at least as fast as $|u'|^{-(D-4)/2}$ for large $u'$. 
Therefore $R$ falls off at least as fast as $|u'|^{-(D-2)}$
and $R_\perp$ falls off at least as fast as
$|u'|^{-(D-3)}$ for large $|u'|$.

Before concluding this section we shall discuss a subtle point. In odd
dimensions, the expansion
of a massless field $\phi_a$ in inverse powers of $r'$ also contains integer 
powers of the form
$r^{\prime -(D-3)}$. In five dimensions this could upset the expansion \refb{eid3s2},
by producing a term  of order $r^{\prime -7/2}$ from the product of
the leading term in $\phi_a$ 
of order $r^{\prime -3/2}$ and the
subleading term of order $r^{\prime -2}$. This is larger
than the subleading term of order $r^{\prime -4}$ given in \refb{eid3s2}. It was shown however
in \cite{1901.05942} that the order $r^{\prime -(D-3)}$ term in the expansion 
of $\phi_a$ is $u'$-independent,
and therefore when we try to construct the stress tensor from the field, we must 
necessarily act either a radial derivative or a transverse derivative on this 
component.\footnote{Since in our set up the sources of the gravitational field
travel with finite velocity at late time, we expect the Coulomb part to be not
$u'$ independent, but it should fall off for $|u'|>> r'$ since the sources move away to
a distance further than $r'$ for $|u'|>> r'$. Nevertheless the important point
is that the $u'$ derivative is of the same order as $r'$ derivative and therefore does
not produce any contribution to $\p_\mu\phi_a$ of order $r^{\prime -2}$.}
This produces an extra power of $1/r'$, making the contribution to the stress 
tensor of order $r^{
\prime-9/2}$.
This is smaller than the subleading term in \refb{eid3s2} which falls off as 
$r^{\prime -4}$.

\subsection{Computation of $\tilde e_{\alpha\beta}$}
 
We shall now substitute \refb{eid3s2} into \refb{emain} to compute $\tilde e_{\alpha\beta}$. 
As usual we divide the integration range into two parts, $r'>L$ and $r'\le L$, to compute the 
contributions to $\tilde e^1_{\alpha\beta}$ and $\tilde e^2_{\alpha\beta}$. We begin with the contribution 
to $\tilde e^1_{\alpha\beta}$, given by
\be \label{ee1rad}
 i\, \NN\, e^{i \, \omega\, |\vec x|} \, \int_{r'>L} d^D x' \, e^{ik.x'}\, T_{R\alpha\beta}(x')
+\hbox{boundary terms at $\infty$}\, .
\ee

First we note that the contribution from the last term in \refb{eid3s2} proportional to
$\p_{u'} N_{\alpha\beta}$ may be analyzed by integration by parts. There are no boundary terms since for
fixed $r'$, integration over $u'\equiv t'-r'$ runs from $-\infty$ to $\infty$ and the integrand falls off at the two ends.
Acting on the $e^{ik.x'}$ factor at fixed $r'$, $\hat n'$,
the $\p_{u'}$ term brings down a factor proportional to $\omega=|k^0|$. 
The integrand multiplying it falls off as $dr'/r'$ for large $r'$, and the $e^{ik.x'}$ factor renders the integral
finite, with at most a $\ln \omega$ divergence for small $\omega$. Since $\omega\ln \omega$ terms are
subsubleading in the soft expansion, we can ignore this term in our computation.

Similarly one can show that higher order terms in the expansion of $T_{\mu\nu}$, beyond those given in
\refb{eid3s2}, do not contribute to \refb{ee1rad} to subleading order. For this let us consider
a term in $T_{\mu\nu}$ of order $r^{\prime -(D-1+a)}$ for any positive number $a$. We substitute this
into \refb{ee1rad} and evaluate it in the $k\to 0$ limit. First let us assume that the integrand falls off
sufficiently fast for large $|u'|$ so that the
$u'$ integral gives a finite result. Then
the integration over the spatial coordinates is proportional to 
$\int_{r'>L}d^{D-1}r'/r^{\prime (D-1+a)}$. This goes as
$L^{-a}$ and is therefore suppressed in the large $L$ limit. Exceptions to this are terms coming from
products of Coulomb components of the fields, which remain $u'$ independent\cite{1901.05942} 
over a time scale of
order $r'$.  After this period the sources producing the Coulomb field will
move away to a distance farther than
$r'$ and the field will begin to decrease. 
Therefore for the contribution to the stress tensor from the product of
these terms, the $u'$ integration can give terms of order $r'$.
However since the Coulomb component appears at order $r^{\prime -(D-3)}$ and its derivatives are of order
$r'^{-(D-2)}$, its contribution to the stress tensor will be of order $r^{\prime -2(D-2)}$. Therefore even if the
$u'$ integral produces a factor of $r'$, the spatial integral will be of the form 
$\int_{r'>L}d^{D-1}r'/r^{\prime (2D-3)}\sim L^{-(D-4)}$. For $D>4$, this is suppressed for large $L$.

Therefore we need to evaluate \refb{ee1rad} 
with $T_{R\mu\nu}$ given by \refb{eid3s2}, ignoring the term proportional to $\p_{u'}N_{\mu\nu}$.  
For this we first make a change of variables:
\be 
\vec x^{\,\prime} \to \vec x^{\,\prime} + {\vec R_\perp(\hat n', u')}/R(\hat n', u')\, .
\ee
This induces the transformations:
\be
r'\to r', \quad u'\to u', \quad  \hat n'\to \hat n' +{\vec R_\perp(\hat n', u')\over r'\, R(\hat n', u')}, 
\quad  d^Dx'\to d^D x' \left\{1 + {1\over r'}\,
\vec \p_\perp . \left({\vec R_\perp(\hat n', u')\over R(\hat n', u')}\right)\right\}\, ,
\ee 
where we have ignored terms that are suppressed by two powers of $r'$.
Using this in \refb{ee1rad} 
we get
\be 
\tilde e^1_{\alpha\beta}=i\,  \NN\,  e^{i\omega |\vec x|}\, \int_{r'>L} 
d^D x'\,  e^{ik.n' \, r' + i \omega u' + i k.R_\perp/R}\, 
\left[{1\over r^{\prime D-2}}\, 
R(\hat n', u')  \,  n'_\alpha \, n'_\beta  + \OO\left({1\over r^{\prime D}}\right)\right]\, .
\ee
In particular the order $1/r^{\prime (D-1)}$ terms in \refb{eid3s2} get cancelled (except for the 
$\p_{u'} N_{\alpha\beta}$ term, which we have argued does not contribute to $\tilde e^1_{\alpha\beta}$  
to subleading
order).
We now express $e^{ik.n' r'}$ as $(ik.n')^{-1} \, \p_{r'}\left( e^{ik.n' r'}\right)$, 
and then integrate by parts over
$r'$, arriving at
\ben\label{est3}
\tilde e^1_{\alpha\beta}&=& - \NN\,  e^{i\omega |\vec x|}\, \int d\hat n' \int du' \, {1\over k.n'}\,
 e^{ik.n' \, L + i \omega u' + i k.R_\perp/R}\, \left[
R(\hat n', u')  \,  n'_\alpha \, n'_\beta  + \OO\left({1\over L^{2}}\right)\right]\nonumber \\ &&
\hskip -.3in 
- \NN\,  e^{i\omega |\vec x|}\, \int_{r'>L} d^D x' \, {1\over k.n'}\,  e^{ik.n' \, r' + i \omega u' + i k.R_\perp/R}\, {\p\over \p r'}
\left[R(\hat n', u')  \,  n'_\alpha \, n'_\beta  + \OO\left({1\over r^{\prime 2}}\right)\right]
\, ,
\een
where $d\hat n'$ denotes integration over the angular variables and we have used:
\be
d^D x' = d \hat n'\, d u'\, r^{\prime D-2} \, dr'\, .
\ee
The first term on the right hand side of \refb{est3} represents the boundary contribution from $r'=L$.
As usual, we have ignored boundary terms at infinity. The term in the second line has integrand of order
$1/(r')^3$, and even when we expand the exponential factor to order $k^\mu$ to pick the subleading term,
the integrand will be of order $1/(r')^2$. Therefore the integral goes as $1/L$ and can be ignored.
This gives, to subleading order in the expansion in powers of $\omega\equiv |k^0|$,
\be\label{enew10}
\tilde e^1_{\alpha\beta} = - \NN\,  e^{i\omega |\vec x|}\,\int d\hat n' \int du' \,  \Bigg[{1\over k.n'}\,
\left\{1+ik.n' \, L + i \omega u' + i k.R_\perp/R\right\}\, 
R(\hat n', u')  \,  n'_\alpha \, n'_\beta 
\Bigg]\, .
\ee

We now turn to the computation of
\be
\tilde e^2_{\alpha\beta}(x) \equiv  i\, \NN\, e^{i \, \omega\, |\vec x|} \, \int_{r'\le L}
d^Dx' \, e^{ik.x'}\, T_{R\alpha\beta}(x')\, .
\ee
The calculation will follow the same steps as the ones described below \refb{eintan}.
We have
\ben\label{enew2}
k_\alpha \tilde e^{2\alpha\beta}(x) &=&  \NN\, e^{i \, \omega\, |\vec x|} \, \int_{r'\le L}
d^Dx' \, \left\{{\p\over \p x^{\prime\alpha}} e^{ik.x'}\right\}\, T_R^{\alpha\beta}(x')\nonumber \\
&=&  \NN\, e^{i \, \omega\, |\vec x|} \, 
\int d\hat n' \int du' \, (r')^{D-2} \, \hat n'_\alpha \, e^{ik.x'}\, T_R^{R\alpha\beta}(x')\bigg|_{r'=L}\, ,
\een
where in the second step we have carried out an integration by parts, picking up the boundary term
at $r'=R$ and using the conservation law\footnote{Actually it is the sum of the 
stress tensor of the radiation and matter that is conserved. So
we really need 
to combine \refb{enew2} with \refb{e2.18} and set to zero the total contribution to
$\p_\mu T^{\mu\nu}$. Similarly
neither \refb{emattcons}
nor \refb{etmncons} is true individually, but their sum is true, and one should analyze the
contribution to $\tilde e^2_{\alpha\beta}$ from matter and radiation together. 
A similar remark holds for the
angular momentum
conservation laws \refb{einr1} and \refb{eradang}. \label{fo1}} $\p_\alpha T_R^{\alpha\beta}(x')=0$.
Since the total ingoing momentum flux is equal to the total outgoing momentum flux, we 
have$^{\ref{fo1}}$
\be\label{etmncons}
- (r')^{D-2} \,\int d\hat n' \int du' \, \hat n'_\alpha \,  T_R^{\alpha\beta}(x')\bigg|_{r'=L} = 0\, .
\ee
Using this we can express \refb{enew2} as
\be\label{enew2newpre}
k_\alpha \tilde e^{2\alpha\beta}(x) =  i\, \NN\, e^{i \, \omega\, |\vec x|} \, (r')^{D-2} \, 
\int d\hat n' \int du' \, \hat n'_\alpha \, k.x'\, T_R^{\alpha\beta}(x')\bigg|_{r'=L} +\OO(\omega^2)\, .
\ee
We can take the solution to \refb{enew2newpre} to be
\be \label{enew2new}
\tilde e^{2\alpha\beta}(x) =
i\, \NN\, e^{i \, \omega\, |\vec x|} \, (r')^{D-2} \, 
\int d\hat n' \int du' \, \hat n'_\gamma \, x^{\prime\alpha}\, T_R^{\gamma\beta}(x')
\bigg|_{r'=L}
+\OO(k)\, .
\ee
This does not look symmetric under $\alpha\leftrightarrow\beta$, but using the
conservation of total angular momentum (see footnote \ref{fo1}):
\be\label{eradang}
(r')^{D-2} \, \int d\hat n' \int du' \, \hat n'_\gamma \, \left[-x^{\prime\alpha}\, T_R^{\gamma\beta}(x')
+ x^{\prime\beta}\, T_R^{\gamma\alpha}(x')\right]\bigg|_{r'=L} = 0\, ,
\ee
we can rewrite \refb{enew2new} as a manifestly symmetric tensor:
\be \label{enew3}
\tilde e^{2\alpha\beta}(x) = {1\over 2}
i\, \NN\, e^{i \, \omega\, |\vec x|} \, 
\int d\hat n' \int du' \, (r')^{D-2} \, \hat n'_\gamma \, \left[x^{\prime\alpha}\, T_R^{\gamma\beta}(x')
+ x^{\prime\beta}\, T_R^{\gamma\alpha}(x')\right]\bigg|_{r'=L}
+\OO(k)\, .
\ee

Adding \refb{enew10} and \refb{enew3}, and dropping terms containing inverse powers of $L$, we now 
get, up to subleading order in the expansion in powers of $k$,
\ben \label{enew4}
\tilde e^{\alpha\beta}(x) &=& \tilde e^{1\alpha\beta}(x)+ \tilde e^{2\alpha\beta}(x)\nonumber \\
&=& - \NN\,  e^{i\omega |\vec x|}\,\int d\hat n' \int du' \,  \Bigg[{1\over k.n'}\,
\big(1+ik.n' \, L + i \omega u' + i k.R_\perp/R\big)\, 
R(\hat n', u')  \,  n^{\prime\alpha} \, n^{\prime\beta} 
\nonumber \\ && \hskip .2in  
- {i\over 2} \, \left[(r')^{D-2}\, \tilde n'_\gamma \, \left\{x^{\prime\alpha}\, T_R^{\gamma\beta}(x')
+ x^{\prime\beta}\, T_R^{\gamma\alpha}(x')\right\}\right]_{r'=L}
 \Bigg]\, . 
\een
We now use the expression for $T_{R\mu\nu}$ given in \refb{eid3s2} to evaluate this expression,
ignoring terms containing inverse powers of $L$. 
Using the result $x^{\prime\mu}=L\, n^{\prime \mu} + u\, (n^{\prime\mu} -\tilde n^{\prime\mu})$ at 
$r'=L$, and 
after an integration by parts in the angular variables for the term proportional to $\vec\p_\perp.\vec R_\perp$,
we get the result:
\be \label{e330fins2}
\tilde e_{\alpha\beta} = -\NN\, e^{i \, \omega\, |\vec x|} \, \int du'\, \int d\hat n'\, 
\Bigg[ R(\hat n', u') \, {n'_\alpha \, n'_\beta\over n'.k} 
-i\, {1\over n'.k}\, k^\gamma\, n'_{(\alpha}
\left\{ \wt R_{\beta)}(\hat n', u') \, n'_\gamma - n'_{\beta)} \, \wt R_{\gamma}(\hat n', u')
\right\}
\Bigg]\, ,
\ee
where
\be \label{ercons2s2}
\wt R_\mu(\hat n', u')= -u' \, R(\hat n', u')\, \tilde n'_\mu + R_{\perp\mu}(\hat n', u')\, .
\ee
It is straightforward to verify that if we substitute the expression for $T_{R\mu\nu}$ given in \refb{eid3s2}
into \refb{eabexpint}, and substitute the resulting expressions for $A_\alpha$, $B_{\beta\gamma}$ into
\refb{esoftfieldint},  we get back the same expression for $\tilde e_{\alpha\beta}$ as given 
in \refb{e330fins2}.\footnote{In the computation we 
have not included the contribution from the
term proportional to $\p_{u'}N_{\alpha\beta}$ in \refb{eid3s2}. This gives a contribution to 
$x^{\prime\alpha}T_R^{\mu\beta} - x^{\prime\beta}T_R^{\mu\alpha} $
proportional to $r^{\prime -(D-2)} \p_{u'} (n^{\prime\alpha} N^{\mu\beta} - n^{\prime\beta} N^{\mu\alpha} )$. Being
a total derivative in. $u'$ (and hence $t'$), its contribution to the soft theorem via \refb{eabexpint} vanishes.
}
This establishes \refb{esoftfieldint}, \refb{eabexpint}.

From the analysis in the penultimate paragraph of \S\ref{s2.1} we know that
$R$ falls off as $|u'|^{-(D-2)}$
and $R_\perp$ fall off as
$|u'|^{-(D-3)}$ for large $u'$. Using these results in \refb{e330fins2}, \refb{ercons2s2} we see that
the integrand in \refb{e330fins2} falls off at least as fast as $|u'|^{-(D-3)}$ for large $|u'|$. Therefore
its integral over $u'$ yields finite result for $D>4$. This observation is particularly relevant for odd $D$ since
the retarded Green's function $G_r(x,x')$ has support for $x$ lying inside the future light-cone of $x'$, instead
of on the future light-cone of $x'$.

\bigskip

{\bf Acknowledgement:}
We would like to thank Miguel Campiglia, Arnab Priya Saha and Biswajit Sahoo for useful discussions.
The work of A.S. was
supported in part by the 
J. C. Bose fellowship of 
the Department of Science and Technology, India and the Infosys chair professorship.

\appendix

\sectiono{Radiative fields at large retarded time} \label{sa}

In this appendix we shall study the asymptotic fall-off of massless fields in the scattering process
at large retarded time. 
We shall assume that we have chosen a
gauge such that the field equation of a massless field $\phi_a$ takes the form:
\be\label{eap1}
\p^\mu \p_\mu \phi_a = -j_a
\ee
for some source term $j_a$. In this case the solution is given by:
\be\label{eap2}
\phi_a = -\int d^D x' \, G_r(x, x') \, j_a(x')\, ,
\ee
where $G_r$ is the retarded Green's function. It was shown in \cite{1801.07719} 
(and reviewed in \S\ref{s1}) that for large $|\vec x|$, 
\refb{eap2} takes the form:
\be
\phi_a(t, \vec x) = {1\over 2\pi} \, \int d\omega\, e^{-i\omega \, t} \tilde\phi_a(\omega, \vec x)\, ,
\ee
with,
\be \label{ebig}
\tilde \phi_a(\omega,\vec x) = {i\over 2\omega}\, e^{i\omega |\vec x|}\,
\left({\omega\over 2\pi i |\vec x|}\right)^{(D-2)/2} 
\int d^D x' e^{i\omega (t' - \hat n.\vec x^{\,\prime})} \, j_a(x'), \qquad \hat n = \vec x/|\vec x| \, .
\ee
Now we know from the analysis of \S\ref{s1}  that for small $\omega$:
\be
\int d^D x' e^{i\omega (t' - \hat n.\vec x^{\,\prime})} \, j_a(x')\simeq {A(\hat n)\over \omega} +\OO(1)\, ,
\ee
for some function $A(\hat n)$. Since small $\omega$ behaviour of $\tilde \phi_a(\omega,\vec x)$ controls the
large time behaviour of $\phi_a(t, \vec x)$, we get, from \refb{eap2}-\refb{ebig},
\be
\phi_a(t, \vec x) \simeq {i\over 4\pi} \,\left({1\over 2\pi i |\vec x|}\right)^{(D-2)/2} \, A(\hat n)\, 
\int d\omega \, e^{-i\omega u} \, \omega^{(D-6)/2}, \quad u\equiv t-|\vec x|\, .
\ee
In even dimensions $>4$ the integral gives $\delta(u)$ or its derivatives\cite{1612.03290,1702.00095}, 
and therefore the expression
is localized around $u=0$. In odd dimensions, 
changing integration variable from $\omega$ to $y\equiv \omega \, u$, we get:
\be
\phi_a(t, \vec x) \simeq {i\over 4\pi} \,\left({1\over 2\pi i |\vec x|}\right)^{(D-2)/2} \, A(\hat n)\, u^{-(D-4)/2}
\int dy \, e^{-iy} \, y^{(D-6)/2}.
\ee
This shows that for $D>4$, $\phi_a(t, \vec x)$ falls off as $u^{-(D-4)/2}$
 for large $u$. This agrees with the results of \cite{1712.00873} and is 
 one of the results used in our analysis in \S\ref{sfield}
for computing the contribution to $\tilde e_{\alpha\beta}$ due to stress tensor of massless fields.

\sectiono{Stress tensor of radiation at large distance} \label{sb}

In this appendix we shall verify the general form \refb{eid3s2} of the stress tensor associated with massless fields
by explicitly constructing the stress tensor of massless scalar, vector and tensor fields.
The asymptotic form of various massless fields that we shall use for this
computation can be found in \cite{1712.01204,1901.05942}, 
but we 
also review their derivation. To simplify notation, we shall drop the subscript $R$ from
$T_{R\mu\nu}$ and drop the primes from the coordinate labels used in \refb{eid3s2}.

We shall work with Bondi coordinates defined as
\be 
u\equiv t-r, \quad r, \quad \theta_K\equiv \hat n_K\, .
\ee
For a given vector $A_\mu$, we get, using \refb{e3.3s2}
\be
A_\mu = \p_\mu u \, A_u + \p_\mu r \, A_r + \p_\mu \theta^K \, A_K
= - n_\mu \, A_u + \tilde n_\mu \, A_r  + {1\over r} \, \p_{\perp \mu} \theta^K A_K\, ,
\ee
so that we have
\be\label{eb3}
A_r = n^\mu \, A_\mu, \quad A_u = (n^\mu - \tilde n^\mu) A_\mu.
\ee
In this coordinate system, the expected form of $T_{\mu\nu}$ given in \refb{eid3s2}
takes the form:
\ben\label{etform}
&& T_{uu} = {1\over r^{D-2}}\, R -{1\over r^{D-1}} \, D^K R_K,
\quad T_{ru}=0, \quad T_{rr}=0\, ,
\nonumber \\ &&
{1\over r} \,  T_{uK} = {1\over r^{D-1}}\, 
R_K, \quad {1\over r}   T_{r K} = 0, \quad{1\over r^2} 
 T_{KL}=0\, .
\een
up to terms that are either of order $1/r^{D-1}$ and total derivative in $u$ or of order $1/r^{D}$.
$R_K$ is related to $R_{\perp i}$ in  \refb{eid3s2}
via the relation,
\be
\p_{\perp i} \theta^K \, R_K = R_{\perp i}\, .
\ee
The metric in this coordinate system is given by
\be \label{ebondigmn}
ds^2 = -2dr\, du - du^2 + r^2 \, Q_{KL} d\theta^K d\theta^L\, ,
\ee
where $Q_{KL}$ is the metric on the unit sphere. The inverse metric
has the form:
\be\label{ebondimet}
g^{ru}=-1, \quad g^{rr}=1, \quad g^{uu}=0, \quad g^{KL} = r^{-2} \, Q^{KL},
\quad g^{uK}=0, \quad g^{rK}=0\, .
\ee
The non-vanishing  Christoffel symbols of the Minkowski metric in Bondi coordinate system
are:
\be\label{ebondicon}
\Gamma^u_{KL} = r\, Q_{KL}, \quad \Gamma^r_{KL} = - r\, Q_{KL}, \quad
\Gamma^K_{rL}=r^{-1}\, \delta^K_L, \quad \Gamma^K_{LM} = \wt \Gamma^K_{LM}\, ,
\ee
where $\wt\Gamma^K_{LM}$ is the Christoffel symbol on the unit $(D-2)$ sphere labelled
by the coordinates $\theta^K$. We shall denote by 
$D_K$ the covariant derivative on the unit sphere 
computed with the metric $Q_{KL}$ and the Christoffel symbol $\wt\Gamma^K_{LM}$,
and by
$D^K$ the combination $Q^{KL} D_L$.

Our goal will be to verify \refb{etform} for massless fields. 
Let us first consider the case of a massless scalar field with stress tensor:
\be\label{efield}
T^\phi_{\mu\nu} = \p_\mu\phi \p_\nu\phi - {1\over 2} \eta_{\mu\nu} \, \p^\rho\phi \p_\rho\phi\, .
\ee
In the Bondi coordinates the Laplace equation $\square \phi = 0$ takes the form:
\be \label{elaplacephi}
-2\p_u \p_r \phi + \p_r^2\phi + r^{-2} D^K D_K \phi 
- r^{-1} (D-2) \p_u\phi + r^{-1} (D-2) \p_r\phi=0\, .
\ee
The 
asymptotic form of the scalar field produced during a classical scattering process has the 
form:\footnote{In standard notation in general relativity, {\it e.g.} in \cite{1901.05942},
$g$ and $\tilde g$ would be denoted as
$\phi^{\left({D-2\over 2}\right)}$ and $\phi^{\left({D\over 2}\right)}$ respectively. We shall avoid using this notation for brevity, but
the translation is straightforward. The same translation can be made for the other fields introduced below,
{\it e.g.} $a_\mu$ and $\tilde a_\mu$ will stand for $\AAA_\mu^{\left({D-2\over 2}\right)}$ 
and $\AAA_\mu^{\left({D\over 2}\right)}$ respectively, and $f_{\mu\nu}$ and $\tilde f_{\mu\nu}$ 
will stand for $h_{\mu\nu}^{\left({D-2\over 2}\right)}$ 
and $h_{\mu\nu}^{\left({D\over 2}\right)}$ respectively.}
\be\label{eformphi}
\phi = {1\over r^{(D-2)/2}} \, g(\hat n, u) + 
{1\over r^{D/2}} \, \tilde g(\hat n, u) + \OO\left({1\over r^{(D+2)/2}} \right) \, .
\ee
Here  $g$ is some function that falls off for large $|u|$ according to the results of
appendix \ref{sa}, 
and $\tilde g$ will be determined 
shortly.  In \refb{eformphi} we have ignored a possible Coulomb term of order
$r^{-2}$ in $D=5$, since we have argued in \S\ref{sfield} that these terms do not
contribute to $T_{\mu\nu}$ to the required order.
Substituting 
\refb{eformphi} into \refb{elaplacephi} we find that the order $r^{-D/2}$ term automatically vanishes and
the order $r^{-(D+2)/2}$ term gives:
\be \label{etg}
\p_u\tilde g(\hat n, u) =- {1\over 2} D^K D_K \, g(\hat n, u) + {(D-2)(D-4)\over 8} \, g(\hat n, u) \, .
\ee
This gives, to order $r^{-D/2}$,
\ben\label{ephder}
&& \p_u\phi =  {1\over r^{(D-2)/2}} \, \p_u \, g(\hat n, u) -
{1\over r^{D/2}} {1\over 2} D^K D_K \, g(\hat n, u) + {1\over r^{D/2}}
{(D-2)(D-4)\over 8} \, g(\hat n, u) \,  ,\nonumber \\
&& \p_r\phi = -{D-2\over 2} \, {1\over r^{D/2}} \, g(\hat n, u), \quad {1\over r} \p_K \phi
= {1\over r^{D/2}} \p_{K} g\, ,
\een
and therefore
\be
\p_\mu\phi \p^\mu\phi = -2 \, \p_u\phi \, \p_r\phi + \p_r\phi\, \p_r\phi 
+ r^{-2}\, Q^{KL}\, \p_{K} \phi\, \p_L\phi 
= {D-2\over r^{D-1}} \, g(\hat n, u )\, \p_u \, g(\hat n, u ) +\OO(r^{-D})\, .
\ee
Since this is a total derivative in $u$, we can ignore its contribution
to $T^\phi_{\mu\nu}$ given in \refb{efield}. From \refb{ephder},
\refb{efield} we now get:
\ben
&& T_{uu} = {1\over r^{D-2}}\, (\p_u g)^2 - {1\over r^{D-1}}\,
\p_u g \, D^K D_K g, \qquad T_{ur}=0\, ,
\qquad
T_{rr}=0,\nonumber \\
&& r^{-1}\, T_{uK}= {1\over r^{D-1}} \, \p_u g\, \p_{K} g,
\qquad  r^{-1} \, T_{rK}=0, \qquad r^{-2}\, T_{KL}=0\, ,
\een
up to terms that are either  of order $r^{-D}$ or of order 
$r^{-(D-1)}$ and total derivative in $u$.
This matches
\refb{etform} up to terms of the form $r^{-(D-1)} \p_u N_{\alpha\beta}$ if we choose:
\be
R= \left( \p_u g \right)^2, \quad
R_K = \p_u g \,  \p_K g\, .
\ee

Next we shall analyze the stress tensor corresponding to the
asymptotic electromagnetic field.
We shall use Lorentz gauge. In the Bondi coordinates, the equations of motion
$\square \AAA_\mu=0$ take the form\cite{1712.01204,1901.05942}:
\ben \label{elaplaceA}
&& \hskip -.2in -2\p_u \p_r \AAA_u + \p_r^2\AAA_u + r^{-2} D^K D_K \AAA_u 
- r^{-1} (D-2) \, \p_u \AAA_u + r^{-1} (D-2)\, \p_r \AAA_u=0\, , \nonumber \\
&& \hskip -.2in  -2\p_u \p_r \AAA_r + \p_r^2\AAA_r + r^{-2} D^K D_K \AAA_r 
- r^{-1} (D-2) \, \p_u \AAA_r + r^{-1} (D-2)\, \p_r \AAA_r 
\nonumber \\  && \hskip 1in
-2\, r^{-3} D^K \AAA_K  + r^{-2} (D-2) \AAA_u - r^{-2} (D-2) 
\AAA_r =0\, , \nonumber \\
&& \hskip -.2in  -2\p_u \p_r \AAA_L + \p_r^2\AAA_L + r^{-2} D^K D_K \AAA_L 
- r^{-1} (D-4) \, \p_u \AAA_L + r^{-1} (D-4)\, \p_r \AAA_L \nonumber \\ &&
\hskip 1in  - 2\, r^{-1} \p_L \AAA_u + 2\, r^{-1} \p_L \AAA_r - (D-3)\, r^{-2} \AAA_L =0\, .
\een
We now express the radiative part of the gauge field in the far region as 
 \be \label{extr1}
 \AAA_\mu(x) = {1\over r^{(D-2)/2}} \, \vea_\mu(\hat n, u) + {1\over r^{D/2}}
 \, \tilde\vea_\mu(\hat n, u)  + \OO\left( r^{(-D-2)/2}\right)\, ,
 \ee
 where the function $\vea_\mu(\hat n, u)$ falls off for large $|u|$. As described
 in \S\ref{sfield}, there are also Coulombic modes\cite{1901.05942}, but their
 contribution to the stress tensor can be ignored at this order.
 The Lorentz gauge condition $\p^\mu \AAA_\mu=0$ gives $\p_u (n^\mu a_\mu)=0$. 
 Since $a_\mu$ falls off at large $|u|$, we get $n^\mu a_\mu=0$. 
 We can use the residual gauge freedom $\AAA_\mu\to \AAA_\mu +\p_\mu\phi$ with
$\square\phi=0$ to also set $\tilde n^\mu a_\mu = 0$. In the $u,r,\theta^K$ coordinate system this gives, from \refb{eb3},
\be\label{egaugeform}
\AAA_u = {1\over r^{D/2}} \tilde a_u, \quad \AAA_r ={1\over r^{D/2}} \tilde a_r,
\quad {1\over r} \AAA_K = {1\over r^{(D-2)/2}} a_K + {1\over r^{D/2}}\tilde a_K\, ,
\ee
up to corrections of order $r^{-D/2-1}$.
 Substituting
 \refb{egaugeform} into \refb{elaplaceA} we get
 equations analogous to \refb{etg}:
 \be \label{eta}
\p_u \tilde a_u = 0,\quad 
\p_u\tilde a_r=D^K a_K, \quad 
\p_u\, \tilde a_{K} = -{1\over 2} D^LD_L a_{K} 
+ {1\over 8} (D^2-6D+12) \, a_{K}\, .
\ee
The first equation, together with the fact that $\tilde a_u$ vanishes in the far
past, allows us to set $\tilde a_u$ to 0.

In the $(u,r,\theta^K)$ coordinate system, different components of the 
field strength $\FF_{\mu\nu}\equiv (\p_\mu \AAA_\nu - \p_\nu \AAA_\mu)$ 
 up to order $r^{-D/2}$ are given by:
 \ben
 && \FF_{ur} = {1\over r^{D/2}} D^K a_K, \quad {1\over r} \FF_{uK} = {1\over r^{(D-2)/2}}
 \p_u a_K + {1\over r^{D/2}} \left\{-{1\over 2} D^L D_L a_K + {D^2 - 6D +12\over 8}a_K
 \right\}\, , \nonumber \\
&&{1\over r} \FF_{rK} = {D-4\over 2} \, {1\over r^{D/2}}
 a_K, \quad {1\over r^2} \FF_{KL} = {1\over r^{D/2}} (D_K a_L - D_L a_K)\, .
 \een
 From this we can calculate the energy momentum tensor:
 \be \label{e311}
 T_{\mu\nu}^{em} = \FF_{\mu\rho} \FF_{\nu}^{~\rho} -{1\over 4} \, \eta_{\mu\nu} \, \FF_{\rho\sigma}
 \, \FF^{\rho\sigma}\, ,
 \ee
 ignoring corrections of order $r^{-D}$ and total derivatives in $u$ in terms of order 
 $r^{-(D-1)}$.
 We first note that to this order
 \be
 \FF_{\rho\sigma} \FF^{\rho\sigma} = -4\, r^{-2} \, Q^{KL}\, \FF_{uK} \FF_{rL}
 = -2\, {D-4\over r^{D-1}}\, Q_{KL} \, a_K \p_u a_L = - {D-4\over 
 r^{D-1}} \p_u \left(Q^{KL} a_K a_L
 \right)\, .
 \ee
 Since  this is a total derivative we can ignore its contribution to $T_{\mu\nu}$.
 Therefore we get, ignoring terms of order $r^{-D}$ and total $u$-derivative terms of
 order $r^{-(D-1)}$:
 \ben
 && T_{uu} = r^{-2} \, Q^{KL}\, \FF_{uK} \, \FF_{uL} = {1\over r^{D-2}} Q^{KL}
 \p_u a_K \p_u a_L - {1\over r^{D-1}} Q^{KL} \, \p_u a_K \, D^M D_M a_L,
 \quad T_{rr} 
= 0\, ,
 \nonumber \\
&& {1\over r} T_{rK} =0, \quad 
T_{ur} = r^{-2} \, Q^{KL}\, \FF_{uK} \FF_{rL } = 0, \quad {1\over r^2} T_{KL} =  {1\over r^2}
g^{ru} (\FF_{Ku} \FF_{Lr} + \FF_{Kr} \FF_{Lu})= 0\, , \nonumber \\ &&
{1\over r} T_{uK} = {1\over r} g^{ru} \FF_{ur} \FF_{Ku} 
+ r^{-2} Q^{LM} {1\over r} \FF_{uL} \FF_{KM} \nonumber \\ && \hskip .45in
= {1\over r^{D-1}}
\p_u a_K \, D^L a_L +  {1\over r^{D-1}} \, Q^{LM} \p_u a_L (D_K a_M - D_M a_K)\, .
\een
This has the same form as \refb{etform} if we identify:
\ben
&& R = Q^{KL}
 \p_u a_K \p_u a_L, \quad R_K = \p_u a_K \, D^L a_L +  
 Q^{LM} \p_u a_L (D_K a_M - D_M a_K), \nonumber \\ &&
 D^K R_K = Q^{KL} \, \p_u a_K \, D^M D_M a_L + \p_u \NN\, .
 \een
 It is easy to check that the last two equations are consistent with each other 
 for suitable choice of $\NN$.
 
 Finally we shall analyze the stress tensor associated with 
 asymptotic gravitational field. We shall use the de Donder gauge:
 \be\label{ededon}
\p^\mu h_{\mu\nu} - {1\over 2} \, \p_\nu h_\rho^{~\rho}=0\, ,
\ee
so that the linearized equations of motion take the form $\square h_{\mu\nu}=0$. 
We expand the radiative part of $h_{\mu\nu}$ in
the far region as
\be 
h_{\mu\nu} = {1\over r^{(D-2)/2}}\, f_{\mu\nu}(\hat n,u) + {1\over r^{D/2}} \, \tilde f_{\mu\nu}(\hat n,u)  
+ \OO\left(r^{-(D-2)/2}\right)\, ,
\ee
ignoring the Coulombic modes as usual.
The gauge condition \refb{ededon} gives, at leading order,
\be\label{ehh1}
n^\mu f_{\mu\nu} -{1\over 2} n_\nu f_\rho^{~\rho}= 0\, .
\ee
In this gauge we still have residual gauge symmetry
\be
h_{\mu\nu}\to h_{\mu\nu} +\p_\mu \xi_\nu + \p_\nu \xi_\mu, \quad \square \xi_\mu =0\, ,
\ee
which induces a transformation
\be
f_{\mu\nu} \to f_{\mu\nu} + n_\mu a_\nu + n_\nu a_\mu\, ,
\ee
for any function $a_\mu(\hat n, u)$. By adjusting $a_\mu$ we can set
\be\label{ehh2}
f_\rho^{~\rho}=0, \quad \tilde n^\rho f_{\rho\mu}=0\, .
\ee
In Bondi coordinates this corresponds to the following expansion of the various components of $h_{\mu\nu}$
up to order $r^{-D/2}$:
\ben\label{ehhexp}
&& h_{uu} ={1\over r^{D/2}}\, \tilde f_{uu}, \quad h_{rr} = {1\over r^{D/2}}\, \tilde f_{rr}, \quad
h_{ru} = {1\over r^{D/2}}\, \tilde f_{ru}, \quad {1\over r}\, h_{uK} = {1\over r^{D/2}}\, 
\tilde f_{uK}, \nonumber \\ &&
{1\over r}\, h_{rK} = {1\over r^{D/2}}\, \tilde f_{rK}, \quad 
{1\over r^2} \, h_{KL} = {1\over r^{(D-2)/2}}\, f_{KL} + {1\over r^{D/2}}\, \tilde f_{KL}, \quad Q^{KL} f_{KL}=0\, .
\een
We can now write down the $\square h_{\mu\nu}=0$ equations in the Bondi coordinate 
system, and
substitute \refb{ehhexp} into these equations to determine $\tilde f_{\mu\nu}$'s in terms of $f_{\mu\nu}$'s as
in the case of scalar fields and gauge fields. 
Explicit form of these equations can be found in \cite{1712.01204,1901.05942}.
For the sake of brevity we shall not describe the full set of
equations for $h_{\mu\nu}$, but give one example. 
The $rr$ component of the equations of motion takes
the form:
\ben
&&-2\p_u \p_r h_{rr} + \p_r^2h_{rr} + r^{-2} D^K D_K h_{rr} 
- r^{-1} (D-2) \, \p_u h_{rr} + r^{-1} (D-2)\, \p_r h_{rr} 
\nonumber \\  &&
-4\, r^{-3} D^K h_{Kr} +2 \, r^{-4} Q^{KL} h_{KL} + 2\, r^{-2} (D-2) (h_{ur}-h_{rr}) =0 \, .
\een
Upon substituting \refb{ehhexp} into this equation we find that the order $r^{-(D+2)/2}$ terms in
the equation
gives
\be
\p_u \tilde f_{rr} = - Q^{KL} \, f_{KL} =0\, ,
\ee
where in the last step we have used the last equation of \refb{ehhexp}. Vanishing of
$\p_u\tilde f_{rr}$ is an important ingredient that was used in \S\ref{sfield} to show that at 
order 
$r^{-(D-1)}$, the transverse component of the gravitational stress tensor is a total
derivative in $u$ -- we shall also see this explicitly in \refb{etmetric}.
Similar analysis with the other components of the $\square h_{\mu\nu}=0$ equation
leads to the following set
of equations for the $\tilde f_{\mu\nu}$'s in the Bondi coordinates:
 \ben\label{eb36}
&& \p_u \tilde f_{uu}=0, \quad \p_u \tilde f_{ur}=0, \quad \p_u \tilde f_{rr}=0, 
\quad \p_u \tilde f_{uK}=0, 
\nonumber \\ &&
\p_u \tilde f_{rK}= D^L f_{KL}, \quad \p_u \tilde f_{KL}= -{1\over 2} D^M D_M f_{KL}
+ {1\over 8} (D^2 -6D+16)\, f_{KL}\, .
\een

We can now use this to compute the energy-momentum tensor of gravitational radiation. In the
asymptotic region we only need to take the terms quadratic in $h_{\mu\nu}$. This is given by\cite{weinberg_book}:\footnote{Even if the action contains higher derivative terms, their contribution to the stress
tensor in the asymptotic region will be suppressed. Therefore we do not include these
terms.}
\be
T_{\mu\kappa}= -h_{\mu\kappa} R^{(1)\rho}_\rho+\eta_{\mu\kappa} h^{\rho\sigma} R^{(1)}_{\rho\sigma}
+ R^{(2)}_{\mu\kappa} -{1\over 2} \eta_{\mu\kappa} \eta^{\rho\sigma} R^{(2)}_{\rho\sigma}\, , 
\ee
where $R^{(1)}_{\rho\sigma}$ and $R^{(2)}_{\rho\sigma}$ represent contributions to the Ricci tensor 
$R_{\rho\sigma}$ linear and quadratic in $h_{\mu\nu}$ respectively:
\be
R^{(1)}_{\mu\kappa}=\p_\mu\p_\kappa h^\rho_\rho -\p_\rho\p_\kappa h^\rho_\mu - \p_\rho\p_\mu
h^\rho_\kappa +\p^\rho \p_\rho h_{\mu\kappa}\, ,
\ee
\ben
R^{(2)}_{\mu\kappa} &=& - 2 h^{\rho\nu} \left[\p_\mu\p_\kappa h_{\rho\nu} -\p_\rho\p_\kappa h_{\mu\nu} 
- \p_\mu\p_\nu h_{\rho\kappa} + \p_\rho\p_\nu h_{\mu\kappa}\right]  \nonumber \\
&& + \left[2 \p_\nu h^\nu_\sigma -\p_\sigma h^\nu_\nu\right] \left[\p_\kappa h^\sigma_\mu +\p_\mu h^\sigma_\kappa
-\p^\sigma h_{\mu\kappa}\right]\nonumber \\ &&
- \left[\p_\rho h_{\sigma\kappa}+\p_\kappa h_{\sigma\rho}-\p_\sigma h_{\rho\kappa}\right]
\left[\p^\rho h^\sigma_\mu +\p_\mu h^{\sigma\rho}-\p^\sigma h^\rho_\mu\right]\, .
\een
In the Bondi coordinates, $T_{\mu\kappa}$ will have the same form, except that the derivatives $\p_\rho$ will
have to be  replaced by $\DD_\rho$ --  covariant derivatives computed with the metric \refb{ebondigmn} and connection
\refb{ebondicon}, and $\eta_{\rho\sigma}$ will have to be replaced by the form of the metric given in 
\refb{ebondigmn}. The calculation is straightforward, yielding the result:
\ben\label{etmetric}
&& T_{uu} = {1\over r^{D-2}} \p_u f_{KL} \p_u f^{KL} -{1\over r^{D-1}} \, \p_u f_{KL} D^MD_M f_{KL}\, ,
\nonumber \\
&& {1\over r} T_{uK} ={1\over r^{D-1}} \, \left[
\p_u f_{MN} D_K f^{MN} - 2\, \p_u f^{LM} D_M f_{LK} + 2 D_M f^{LM} 
\, \p_u f_{LK}\right]\, ,
\een
with all the other components vanishing to this order.
This has the form given in \refb{etform} with:
\ben
&& R = \p_u f_{KL} \p_u f^{KL}, \qquad R_K = \p_u f_{MN} D_K f^{MN}
 - 2\, \p_u f^{LM} D_M f_{LK} + 2 D_M f^{LM} 
\, \p_u f_{LK}, \nonumber \\ &&
D^K R_K = \p_u f_{KL} D^MD_M f_{KL} +\p_u\NN\, .
\een
In particular the last two equations are compatible for suitable choice of $\NN$. The expression
for $T_{uK}$ given in \refb{etmetric} also agrees with the result of \cite{1502.06120} in $D=4$.

\end{document}